\documentclass[10pt,preprint]{aastex} 
 
\newcommand{\teff}{$T_{\rm eff}$}
\newcommand{\logg}{$\log~g$}

\begin{document}
 
\title{A large stellar evolution database for population
synthesis studies. I. Scaled solar models and isochrones}
 
\author{Adriano Pietrinferni\altaffilmark{1}}
\affil{INAF - Osservatorio Astronomico di Teramo, Via M. Maggini,
64100 Teramo, Italy; pietrinferni@te.astro.it}

\author{Santi Cassisi\altaffilmark{2}}
\affil{INAF - Osservatorio Astronomico di Teramo, Via M. Maggini,
64100 Teramo, Italy; cassisi@te.astro.it}
 
\author{Maurizio Salaris}
\affil{Astrophysics Research Institute, Liverpool John Moores University,
Twelve Quays House, Birkenhead, CH41 1LD, UK; ms@astro.livjm.ac.uk}
 
\and
\author{Fiorella Castelli\altaffilmark{3}}
\affil{Istituto di Astrofisica Spaziale e Fisica Cosmica, CNR, Via del Fosso del Cavaliere, 
00133, Roma, Italy; castelli@ts.astro.it}
 
\altaffiltext{1}{Dipartimento di Statistica, Universit\`a di Teramo, Loc. Coste
S. Agostino, 64100 Teramo, Italy}

\altaffiltext{2}{Instituto de Astrof\'isica de Canarias, 38200 La Laguna, Tenerife,
Canary Islands, Spain}

\altaffiltext{3}{INAF - Osservatorio Astronomico di Trieste, via Tiepolo 11, 34131 
Trieste, Italy; castelli@ts.astro.it}

\normalsize
\begin{abstract}
 
\noindent
We present a large and updated stellar evolution database
for low-, intermediate- and high-mass stars  
in a wide metallicity range, suitable for studying 
Galactic and extragalactic simple and
composite stellar populations using population synthesis techniques. 
The stellar mass range is 
between $\sim0.5M_\odot$ and $10M_\odot$ with a fine mass spacing.
The metallicity [Fe/H] comprises 10 values ranging from $-2.27$ to
$0.40$, with a scaled solar metal distribution. 
The initial He mass fraction ranges from Y=0.245, 
for the more metal-poor composition, up to 0.303 for the more
metal-rich one, with $\Delta Y/\Delta Z\sim 1.4$. 
For each adopted chemical composition, the evolutionary models have been
computed without (canonical models) and with overshooting from the
Schwarzschild boundary of the convective cores during the central
H-burning phase. Semiconvection is included in the treatment of core
convection during the He-burning phase.

The whole set of evolutionary models can be used to compute
isochrones in a wide age range, from $\sim30$ Myr to $\sim15$Gyr.
Both evolutionary models and isochrones are available in several observational planes,
employing updated set of bolometric corrections and color-$T_{eff}$
relations computed for this project. The number of points along the
models and the resulting isochrones is selected in such a way that 
interpolation for intermediate metallicities not contained in the grid
is straightforward; a simple quadratic interpolation produces results
of sufficient accuracy for population synthesis applications.

We compare our isochrones with results from a series of widely used 
stellar evolution databases and perform some empirical tests for the
reliability of our models. 
Since this work is devoted to scaled solar
chemical compositions, we focus our attention on the Galactic disk
stellar populations, employing multicolor photometry of 
unevolved field main sequence stars
with precise $Hipparcos$ parallaxes, well-studied open clusters and 
one eclipsing binary system with precise measurements of masses, radii
and [Fe/H] of both components. 
We find that the predicted metallicity dependence of the
location of the lower, unevolved main sequence in the Color Magnitude Diagram
(CMD) appears in satisfactory agreement with empirical data. When comparing our models
with CMDs of selected, well-studied, open clusters, once again we were able to
properly match the whole observed evolutionary sequences by assuming
cluster distance and reddening estimates in satisfactory agreement with empirical evaluations
of these quantities.

In general, models including overshooting during the H-burning phase
provide a better match to the observations, at least for ages below
$\sim$4~Gyr. At [Fe/H] around solar and higher ages (i.e. smaller
convective cores) before the onset of radiative cores,  
the selected efficiency of core overshooting may be too high in our 
as well as in various other models in the literature.
Since we provide also canonical models, the reader is strongly
encouraged to 
always compare the results from both sets in this critical age range.

\end{abstract}


\section{Introduction}

\noindent
The availability of large sets of stellar evolution models 
spanning a wide range of stellar masses and initial chemical
compositions is a necessary
prerequisite for any investigation aimed at interpreting observations
of Galactic and extragalactic, resolved and unresolved  stellar populations.

A reliable library of stellar evolution models used to study stellar populations
in various environments has to satisfy at least three main criteria:
1) the input physics has to be up to date and the treatment of physical
processes has to be adequate and as much accurate as possible;
2) the set of models has to be homogeneous, in the sense that all
evolutionary phases and initial chemical compositions have to be 
computed with the same evolutionary code and the same physical framework; 3) 
the models have to reproduce as many empirical constraints as possible.
 
The accuracy of the results obtained from methods based on stellar
evolution models is strongly dependent on the reliability of the physical
framework adopted in the computations. 
In this respect, it is very interesting to notice that only about a decade ago the
available stellar models were already producing models that matched 
reasonably well several observational
constraints. However, new results from helioseismology came to show that solar models 
available at that time were seriously challenged by these new empirical
results. This stimulated a large effort toward
improving our knowledge of the physics of stellar matter, 
and as a consequence, in these last years, several
aspects of stellar physics (e.g., radiative opacities, equation of
state) have been largely improved. This process has produced a
continuous improvement in solar models as well as in the whole stellar
evolution field, with non-negligible
variations in quantitative predictions, as in the case of the
Galactic globular cluster ages (e.g. Chaboyer et al.~1995; Mazzitelli, D'Antona \&
Caloi~1995; Salaris, Degl'Innocenti \& Weiss~1997).
A detailed discussion of the effect of recent improvements in the
input physics on the stellar evolution results can be found in Cassisi
et al.~(1998). This evidence provides a clear support for the need of a continuos
improvement and update of stellar model databases.

The homogeneity of the adopted model library is also extremely
important, because the use of different sets of models (computed by
different authors with different input phsyics) for 
different evolutionary phases and/or chemical
compositions and/or mass ranges can introduce subtle inconsistencies in the
interpretation of observations, and can affect the predictive power of
stellar population synthesis techniques.

As for the comparison with empirical constraints, it goes without
saying that this is the best way to assess the reliability of the
chosen evolutionary scenario. In particular, it is very 
important to compare theoretical models with
empirical constraints from resolved stellar populations, such as
Galactic field stars and Galactic stellar
clusters. This is a fundamental step in order to be able to estimate
the level of accuracy of the models, as well as for evaluating the uncertainties 
which could affect the population synthesis analysis based on the same models.

The need of continuously improved stellar evolution databases (in spite of the already
existing large number of stellar models and isochrone sets) incorporating the latest advances
in modeling the stellar physics cannot be more emphasized by the
fact that existing differences among stellar model libraries 
translate into large uncertainties in, just to give an example, 
the ages of external galaxies; this is well illustrated by, e.g.,   
the large differences obtained by Spinrad et al.~(1997) and Yi et al.~(2000),
regarding the age (at least 3.5 Gyr vs. 1.4--1.8 Gyr, respectively) 
of an unresolved high-redshift galaxy (LBDS 53W091),
due mainly to large differences in the colors (and spectra) 
predicted by the two different sets of isochrones used in the two
mentioned papers. Such large differences due only to uncertainties in 
the modeling of stellar populations have 
important consequences for our interpretation of the epoch and
mechanisms of galaxy formation.

Only by employing independent sets of stellar model databases one can assess the
real uncertainties in the inferred parameters for resolved and
unresolved stellar populations. In case of some key stellar input
physics, like, i.e., boundary conditions, overshooting,  
atomic diffusion efficiency, mass loss, 
it is not yet clear what the best prescription is; to this one has to
add current uncertainties on the synthetic spectra used to transform the
output of theoretical stellar models into observed quantities like
colors, magnitudes, line strengths. 
It is therefore necessary the availability of 
independent sets of models, computed with the
best possible choices of the well established physics, and with various
choices for the parameters describing the physical processes not yet
well understood.
It is therefore important that several sets of
models are available, computed with the best possible inputs for the well
established  physics (e.g., equation of state, opacities), and with
various choices of the parameters describing the physical processes 
not yet well understood (e.g., mixing, convection, mass loss).

The recent literature is rich of updates of stellar model libraries, 
see, e.g.,  Bono et al. (2000), Girardi et al. (2000), VandenBerg et al. (2000),
Lejeune \& Schaerer~(2001), Yi et al. (2001) 
and Castellani et al. (2003); however not all of
them cover homogeneously a large range of ages and chemical compositions.
Over the years, we have published a lot of specialized papers
with a series of specific goals, other than addressing the issue of
population synthesis techniques. We have investigated
the effects of improvements in the adopted physical inputs on stellar
models (Salaris \& Cassisi~1996; Brocato, Cassisi \& Castellani~1998;
Cassisi et al. 1998, 1999 and references therein); we have compared
our low mass isochrones with selected observational constraints 
from old stars (e.g., 
Cassisi \& Salaris~1997; Salaris \& Cassisi~1998; Riello et al.
2003) and obtained 
estimates of Galactic globular cluster ages (e.g., Salaris,
Degl'Innocenti \& Weiss~1997; Salaris \& Weiss~1998, 1999, 2002), 
initial He abundance of globular cluster stars (Cassisi,
Salaris \& Irwin 2003; Salaris et al. 2004), pulsational properties of
variable stars  (De Santis \& Cassisi
1999; Bono et al. 1997). This notwithstanding, 
we have never published an extended and homogeneous set of
stellar models covering a large range of stellar masses, chemical
compositions and evolutionary phases, 
suitable for population synthesis analysis as well as for any other
kind of investigation making extensive use of stellar evolution models.
 
The main purpose of this work is to fill this gap, 
by providing an up-to-date and complete set of stellar
models for both low-and intermediate mass, and high mass stars up to
10 $M_{\odot}$, spanning a large metallicity regime from 
metal-poor star systems to super-metal-rich populations. 
Our updated theoretical models are then coupled to a new set of
color-transformation and bolometric corrections.
In this paper, all theoretical 
results are based on
a scaled solar metal distribution. 
The case of an $\alpha-$element
enhanced metal mixture will be the subject of a forthcoming paper.

The paper is organized as follows: \S~2 summarizes the improvements made to our
evolutionary code, while \S~3 deals specifically with the
difficult problem of the core convection treatment. Our standard solar
model is briefly discussed in \S~4, and the model library is
presented in \S~5. Comparisons with the most used existing isochrone
databases, and with selected empirical constraints are discussed in
\S~6, and \S~7, respectively. A summary and conclusions follow in \S~8.

\section{The stellar evolution code and input physics}
 
\noindent
The stellar evolution code adopted in this work is the same used by 
Cassisi \& Salaris~(1997) and Salaris \& Cassisi~(1998), with various updates. 
We have added the nuclear burning of light elements as Li, B and
Be, improved the numerical procedures that optimize the choice
of both the number of mesh points within a model, and the evolutionary
time steps, and updated the model input physics. 
In the following we summarize our choice of the most relevant input
physics, 
and -- in case of differences with our previous
Cassisi \& Salaris~(1997) models -- we discuss its impact on relevant
model parameters.

\begin{itemize}
\item
The radiative opacities are from the OPAL tables
(Iglesias \& Rogers~1996) for temperatures
larger than $10^4$ K, whereas the opacities by Alexander \& Ferguson~(1994)
which include the contribution from molecules and grain have been
adopted for lower 
temperatures.  
Both high and low temperature opacity tables assume a scaled solar heavy element 
distribution (Grevesse \& Noels~1993).
As for the electron conduction opacities we use the recent results by 
Potekhin et al.~(1999) and Potekhin~(1999, hereinafter P99).
These computations cover also the physical conditions typical of
degenerate He cores in Red Giant Branch stars; this avoids the
extrapolation one had to perform with our previous choice of the Itoh
et al.~(1983, 1993, hereinafter - I93) conductive opacities since, as first shown by  
Catelan, de Freitas Pacheco \& Horwath~(1996), the I93 opacities  
did not cover the degenerate He cores in low mass stars.
Before computing the whole set of evolutionary
models presented in this work, we have tested the effect of the P99 opacity on the size
of electron degenerate He cores at the He flash ignition: 
at a metallicity Z=0.002 the change of the He core mass at the He flash
in a $0.8M_\odot$ model is equal to $6\times10^{-4}M_\odot$, 
in the sense that the P99 opacities produce a 
less massive He core with respect the I93, but at solar
composition the trend reverses, the He core mass being $\approx
7\times10^{-4}M_\odot$ larger when using the 
P99 opacity.
It is clear that the effect related to the use of the P99 data is not
large. This notwithstanding we decided to update our code by using the P99 
opacities, in order to avoid any extrapolation.

\item
We updated the energy loss rates for plasma-neutrino processes
(relevant in the He degenerate cores) using the most recent and accurate results 
provided by Haft, Raffelt \& Weiss (1994). 
For all other processes we still rely on the same prescriptions adopted by Cassisi \& Salaris~(1997).
In case of a 1$M_{\odot}$ star with $Z$=0.002, these new neutrino
rates produce a He core mass at the
He flash $\sim$0.005 $M_{\odot}$ larger (and the corresponding
$log(L/L_{\odot}$ is increased by $\sim$0.025) than our
previous results based on the Munakata, Kohyama \& Itoh~(1985) 
rates. However, the same 1$M_{\odot}$ star at solar
metallicity (see \S~4) shows an He core mass at the He flash decreased by 0.003
$M_{\odot}$ (the corresponding
$log(L/L_{\odot}$ is decreased by $\sim$0.015) with respect to our
previous models. 

\item
The nuclear reaction rates have been updated by using values from the
NACRE database  (Angulo et al.~1999), with the exception of the
$^{12}$C$(\alpha,\gamma)^{16}$O reaction. For this reaction we employ
the more accurate recent 
determination by Kunz et al. (2002). Electron screening is treated
according to Graboske et al.~(1973).
We have tested the effect on the main sequence turn
off luminosity when passing from our previously adopted rates
(Caughlan \& Fowler~1988) to the NACRE ones. We found that for a
1$M_{\odot}$ star of solar composition the age at the turn off
increases by 0.6 \%, and log($L/L_{\odot})$ at the turn off increases by
$\sim$ 3.5\%.

The new $^{12}$C$(\alpha,\gamma)^{16}$O rate by Kunz et al.~(2002)
reduces the He burning evolutionary lifetimes 
(at a fixed He core core mass and envelope composition) 
by approximately 7-8 \%, with respect to the results obtained with our
previously adopted Caughlan et al.~(1985) rate.

\item 
The detailed Equation of State (EOS) by A. Irwin\footnote{The EOS code
is made publicly available at 
ftp://astroftp.phys.uvic.ca under
the GNU General Public License (GPL)} has been used. A full
description of this EOS is still in 
preparation (Irwin et
al.~2004) but a brief discussion of its main characteristics can
be found in Cassisi, Salaris \& Irwin (2003). 
This EOS can cover the entire stellar structure along all the main
evolutionary phases of stars in a large mass
range (including the full mass range spanned by our present models);
the accuracy of this EOS is similar to the case of the OPAL EOS 
(Rogers, Swenson \& Iglesias~1996) -- and recently updated in the
treatment of some physical inputs (Rogers \& Nayfonov 2002) -- that
however does not allow the full coverage of the stellar models
presented in this work. 
In fact, e.g., the new OPAL EOS has still an upper
limit to the temperature of $10^{8}$ K, and this prevents any
evolutionary computation beyond the approach to the He burning ignition.

To give an idea of the differences between our adopted EOS
and the OPAL one in the common range of validity, we show in
Fig.~\ref{sspeed} a comparison of the run of the ratio U=P/$\rho$ as a
function of the distance from the star center, obtained from
our calibrated solar model (see \S~4), and a solar model calibrated
using the OPAL EOS. In both calibrated models the surface $Z/X$ ratio, envelope He
abundance and depth of the convective envelope match the 
empirical determinations (see below and \S~4).
Besides the comparison with the values of U inferred from
helioseismology (data from Degl'Innocenti et al.~1997) it is important to
notice how our adopted EOS closely matches the results obtained with
the OPAL one.

As for the effect of passing from the EOS used in Cassisi \&
Salaris~(1997) to the actual one, we performed a test by computing the
evolution of a 1$M_{\odot}$ star of solar metallicity up to the He
flash, using both EOSs and keeping everything else fixed.
With the new Irwin EOS the zero age main sequence log$L/L_{\odot}$ is $\sim$25
\% brighter, and its effective temperature $\sim$250 K hotter; the
turn off log$L/L_{\odot}$ is $\sim$10 \% brighter, its effective
temperature $\sim$190 K higher and its age is 1.6 Gyr younger ($\sim$16 \% younger).
At the He flash this age difference is preserved, whereas the He core
mass is 0.01 $M_{\odot}$ smaller.

\item
Our stellar evolution code can account for the atomic diffusion of both Helium 
and heavy elements. The diffusion coefficients are calculated by means of a routine provided
by A. Thoul (1995, private communication), which solves Burgers's equation for a multicomponent
fluid (Thoul, Bahcall \& Loeb~1994, see also the in depth discussion
by Schlattl \& Salaris~2003). At variance with canonical computations
in which the total metallicity does not change within the star, when diffusion
occurs the metal abundances display a gradient across the stellar structure. However,
being all elements heavier than carbon more or less equally diffused,
the abundance ratios are not
significantly affected by atomic diffusion, and this allows to account for the effects
of diffusion on the stellar opacity simply by interpolating among
scaled-solar opacity tables of different Z.

We have properly accounted for atomic diffusion
only when computing the standard solar models (see \S~4). The whole set
of evolutionary models presented in this paper has been computed
without including atomic diffusion. In fact, although in the Sun
atomic diffusion is basically fully efficient, spectroscopic
observations of, e.g., stars in Galactic globular clusters or field halo stars 
(see, e.g., Gratton et al.~2001; Bonifacio et al.~2002 and references therein)
point to a drastically reduced efficiency of diffusion, maybe due to
the counteracting effect of some sort of 
rotational mixing processes. 
This raises the possibility that the almost uninhibited efficiency of
diffusion found in the Sun might not be a common occurrence.
In view of this uncertainty we decided,
conservatively, to not include in our model grid computation the
effect of atomic diffusion. 

\item
The outer boundary conditions have been computed integrating the
atmospheric layers with the $T(\tau)$ relation
provided by Krishna-Swamy~(1966). 
Superadiabatic convection is treated according to the Cox \&
Giuli~(1968) formalism of the mixing length theory (B\"ohm-Vitense~1958).
The mixing length parameter 
has been fixed by the solar calibration and kept constant 
for all masses during all evolutionary phases (see, e.g., Freytag \&
Salaris~1999).

\item
All models include mass loss using the Reimers formula (Reimers 1975) with the free
parameter $\eta$ set to 0.4.
This value of $\eta$ is commonly assumed (e.g. Girardi et al.~2000)
because it allows to broadly reproduce the mean colors of horizontal
branches in Galactic globular clusters.
We are in the process of computing an additional set of models (for all metallicities
in our grid) that employs $\eta$=0.2, in order
to provide a tool to assess the impact of different efficiency of mass
loss on various properties of simple and composite stellar populations
(e.g., their integrated colors).

\end{itemize}


\section{The treatment of convection at the border of convective core.}

\noindent
Stars with masses larger than $\sim 1.1 
M_{\odot}$ (the exact value depends on the initial chemical
composition) during the central H-burning phase develop a 
convective core, due to the dependence of the CNO-cycle efficiency on temperature.
The instability against 
turbulent convection is classically handled by means of the Schwarzschild 
criterion for a chemically homogeneous fluid. This well known
criterion is based on the comparison between the 
expected temperature gradient produced by the radiative transport of
energy, and the adiabatic one. It  is worth noticing 
that the proper estimate of the size of the convective region is then 
primarily dependent on the accuracy of the input 
physics. Any improvement of the adopted physics may produce a change
in the temperature gradients and, 
in turn modify the location of the boundaries of the convective regions. 
 
A second important point concerns the possibility that the motion of
the convective fluid elements is not halted 
at the boundary with the stable region beyond the formal
convective core. Although beyond the Schwarzschild boundary a moving
fluid element is subject to a strong deceleration, it might be possible that a 
nonzero velocity is mantained along a certain length. This mechanical
overshoot may induce a significant amount of mixing in a region
formally stable 
against convection (see Cordier et al.~2002, and references therein).  

Finally, another important 
point which has to be taken into account is that a sizeable increase
in the internal mixed region, hence 
larger convective cores, could be obtained as a consequence of rotationally 
induced mixing (Meynet \& Maeder~2000). 
 
There are therefore at least three different reasons (inadequate input
physics, overshooting from the convective boundary, additional
mixing due to rotation) why the size of the 'real' stellar 
convective cores might be different from the one predicted by 
\lq{canonical}\rq\ models, that is, models computed neglecting
rotation, and overshooting from the Schwarzschild convective boundary.  
This evidence raises the question if the size 
of the convective core, as determined by the (classical) Schwarzschild criterion 
is able to properly match the observations, or it has to be \lq{artificially}\rq\ increased.

Discrepancies with observations have been usually interpreted in terms of
the efficiency of an {\sl overshooting} mechanism;
the extension of this additional mixed
region is usually defined in terms of a 
parameter $\lambda_{OV}$ which gives the length -- expressed as a fraction of
the local pressure scale height $H_P$ -- crossed by the convective 
cells in the convectively stable region outside the Schwarschild convective boundary.  
 
The case for a significant amount of overshooting has been presented
many times both theoretically and 
observationally, although in some cases the results 
have been contradictory (see, e.g., Testa et al.~1999; Barmina, Girardi \&
Chiosi~2002; Brocato et al.~2003). What is well known is the effect of
including overshooting in the stellar models: during the core
H-burning phase the star 
has a larger He 
core, a brighter luminosity and a longer lifetime. During the
following core He-burning  phase, the luminosity is brighter, the
lifetime is shorter and the blue loops in the Color Magnitude Diagram
(CMD) are less extended than in the 
absence of overshooting. Theory therefore predicts that the mass-luminosity (M-L) 
relationship for stars crossing the Cepheid instability strip is
largely 
affected by the amount of overshooting 
accounted for in the stellar evolution computations. The comparison of
the theoretical M-L 
relationship with empirical 
data could, in principle, put tight constraints on the efficiency of
this process. This approach 
has been recently adopted  by
Keller \& Wood (2002) by adopting a large database of empirical data
for Bump Cepheids and, the 
obtained results seem to support the
occurrence of a large overshooting efficiency. Although the method is
sound, the results have to be cautiously treated, as clearly shown by Cassisi (2004).

Due to the lack of a general consensus about the convective core overshooting 
efficiency our stellar models are computed both without overshooting
and with a significant efficiency of this process. In the latter case,
we adopt $\lambda_{OV}=0.20\times{H_P}$. 
This value allows a good match between to the CMD of Galactic open
clusters
of different ages, and allows us a comparison with 
the evolutionary models recently presented by other groups, all
using a similar value of the overshooting parameter.
The overshooting region is fully mixed as far as the chemical elements
are concerned, and the temperature gradient is kept at its radiative value.

Another important issue to be addressed is the value of
$\lambda_{OV}$ when stars have small convective cores.
It is well known that the Turn Off (TO) morphology of the evolutionary tracks and, in turn, of
the isochrones, does depend on how the extent of the convective core
decreases with decreasing mass. When moving 
deeper and deeper inside the star, the pressure scale height steadily
increases; this causes a large increase of the size of convective cores
in stars whose Schwarzschild convective boundary is fast shrinking, 
(e.g., for masses below $\approx1.5 M_{\odot}$)
if the overshooting efficiency is kept fixed at a constant fraction of $H_P$. 
It is clear the need to decrease $\lambda_{OV}$ to zero for stars with
small convective cores, a problem addressed theoretically by, e.g., 
Roxburgh~(1992) and Woo \& Demarque~(2001).

In order to show how critical this issue is, we show in Fig.~\ref{over0} 
two isochrones of the same
chemical composition and age, obtained using different
assumptions about the trend of  $\lambda_{OV}$ with mass.
The change in the isochrone 
morphology is quite significant and different choices concerning the
core overshoot efficiency in the critical mass range $1.0\ge{M/M_\odot}\le1.5$\footnote{It is 
worth noticing that the finer
details of the isochrone morphology do depend not only on the adopted overshoot efficiency, but
also on the number of evolutionary tracks used for computing the isochrones.} mimic
different isochrone ages. 
It is also worthwhile to notice the significant change
in the shape of the TO region. 
This means that the trend of $\lambda_{OV}$ with mass, for masses with
small convective cores, potentially introduces an
additional degree of freedom in stellar evolution models. 
Even if this problem affects only a restricted range of cluster ages, 
it has to be taken into 
account when discussing the uncertainties affecting stellar models.

In our models with core overshooting, regardless of the initial metallicity, we have chosen
the following algorithm for varying the overshoot efficiency with
mass: for masses larger or equal to 1.7$M_\odot$ we fixed
$\lambda_{OV}$ at 
0.20$H_P$; for stars less massive than $1.1M_\odot$ we use
$\lambda_{OV}=0$ (we checked that all our models of at least $1.1M_\odot$ show a
central convective core during the MS), while in the intermediate range
$\lambda_{OV}$ varies according to the relation 
$\lambda_{OV}$=$(M/M_{\odot}-0.9)/4$. We have verified by performing
several numerical experiments, that this choice allows a smooth
variation of the isochrone TO morphology, and a smooth decrease to
zero of the convective cores for stars in this mass range.

Moreover, this choice for the overshooting extension does provide a
good fit to the TO morphology of a sample of clusters of various ages,
as discussed later. In particular, our overshooting isochrones fit
extremely well the TO morphology of clusters like NGC~2420 and
NGC~6819 where the typical TO mass is $\sim 1.3M_{\odot}$ (for
[Fe/H]=$-$0.44) and 
$\sim 1.4M_{\odot}$  (for [Fe/H]=0.06), respectively, i.e. in a mass range where the
$\lambda_{OV}$ is decreasing according to our prescription.
The case of M~67 (Sandquist~2004, see also below) 
and the eclipsing binary AI Phoenicis (see below)
seem however to indicate that $\lambda_{OV}\sim$0 when the stellar mass is
equal to $\sim 1.2M_{\odot}$. Since we provide both overshooting and
canonical models, the user can freely choose to use a criterion
based on age for the switch to $\lambda_{OV}$=0 
models and isochrones.

As for the convective cores during the  He-burning phase, 
we account for semiconvection (which is driven by mechanical
overshooting from the boundary of the Schwarzschild core) at the border
of the canonical convective core by using the numerical scheme
described by Castellani 
et al.~(1985).
Near the core Helium exhaustion we inhibited the breathing pulses (Castellani et al. 1985;
Dorman \& Rood 1993), following the results by Cassisi et al.~(2003). 
The breathing pulses are suppressed adopting the procedure suggested by Caputo et
al.~(1989), i.e., by limiting the extent of the convective core so
that the 
central Helium abundance 
cannot increase between consecutive models. 

We neglect the possible
occurrence of overshooting from the bottom of the convective envelopes. 
This choice has been made in order to reduce as much as
possible the number of free parameters in the stellar computations,
and also in light of the fact that there is no clear-cut empirical
evidence for a sizable efficiency of this phenomenon. A critical test
for this phenomenon is the comparison between the position of the RGB
bump in Galactic globular clusters with the theoretical counterpart. A
non negligible amount of overshooting from the bottom of the
convective envelope does shift the bump to lower luminosities. 
This kind of test does not provide a definitive answer 
(e.g. Riello et al.~2003), mainly because of the uncertainty on the
metallicity scale of globular clusters (the bump position is also
strongly affected by the initial stellar metallicity).
It is however important to remark 
that convective envelope overshooting can affect the morphology of blue loops during the core
He-burning phase (Alongi et al. 1991; Renzini et al. 1992;
Renzini \& Ritossa 1994). A detailed investigation of the dependence of blue loops
on the physical assumptions adopted in stellar computations and a
comparison with Cepheid data will be
the subject of a forthcoming paper.


\section{The standard solar model}

\noindent
The calibration of the standard solar model allows one to fix the value of the mixing length,
as well as the solar initial He and metal fractions. In practice, one
finds the combination of initial He ($Y_{\odot}$) and  metal ($Z_{\odot}$) mass
fractions, and the value of the mixing length parameter that reproduce the
solar radius ($R_{\odot}=6.960\times10^{10} cm$),
luminosity ($L_\odot=3.842\times10^{33} ergs~s^{-1}$, Bahcall \& Pinsonneault 1995) and
ratio $(Z/X)=0.0245\pm0.005$
(Grevesse \& Noels~1993), at an age of $t_\odot=4.57$~Gyr.

The accuracy of the theoretical solar model can be then tested by
comparing the depth of the outer convective layers and the surface $Y$ value
with the values 0.710 -- 0.716$R_\odot$ (Castellani et al. 1997),
and  0.2438 and 0.2443 (Dziembowski et al.~1995), respectively, as
obtained from helioseismological studies. 

It is well known 
(see, e.g., Bahcall \& Pinsonneault 1992, 1995; Basu et al. 1996; Bahcall,
Pinsonneault \& Basu 2001 and references therein) that solar
models without microscopic diffusion cannot properly account for some
features of the seismic Sun. For this reason, we decided to account for atomic 
diffusion of both Helium and metals when computing solar models, as described in section~2.

We have computed several models with mass equal to $1M_\odot$ starting from the
pre-MS phase, under different assumptions about the initial He and metal abundances, and mixing
length parameter. 
The \lq{best}\rq\ solar model which fulfills the quoted requirements
has been obtained for an initial $Y_{\odot}$=0.2734, an initial
$Z_\odot=0.0198$, and a mixing
length value equal to 1.913. This standard solar model is in agreement with the
results from helioseismology concerning the depth of 
the convective envelope ($0.716R_\odot$) and
the present surface He abundance ($Y=0.244$), and with the actual  
(Z/X) ratio ((Z/X)=0.0244). 
The value of the mixing length obtained from the 
calibration of the solar standard model is then 
adopted for the whole set of models presented in this work. We will
show later that this choice of the mixing length coupled with the
color-effective temperature transformations
used in this work allows a fine match to several observational constraints. 
The initial solar He abundance and metallicity ($Y_\odot=0.2734 -
Z_\odot=0.0198$) 
are the values
adopted when computing the so called 'solar metallicity' model set. 

\section{The stellar model library}

	
\subsection{The initial chemical composition of the model grid}

\noindent
As already stated, this work does provide an homogeneous and
self-consistent database of theoretical stellar 
models and isochrones,
based on the scaled solar heavy element distribution by Grevesse \& Noels (1993).
The extension to models accounting for a suitable enhancement of $\alpha$-elements
will be the subject of a forthcoming paper.

In order to cover a wide range of chemical compositions, we provide models 
computed for 10 different metallicities, namely $Z=0.0001$, 0.0003,
0.001, 0.002, 0.004, 0.008, 0.01, 0.0198, 0.03 and 0.04. 

As far as the initial He-abundances are concerned,
we adopt for the lowest metallicity the estimate recently provided by Cassisi et al. 
(2003, see also Salaris et al. 2004) based on 
new measurements of the $R$ parameter in a large sample of Galactic
globular clusters. 
They found an initial He-abundance for globular cluster stars of 
the order of $Y=0.245$, which is in satisfactory agreement with the recent
determination of the cosmological 
baryon density provided 
by WMAP observations of the power spectrum of the Cosmic Microwave
Background (CMB) radiation (Spergel et al.~2003).
It is worth noticing that the theoretical $R$ parameter
calibration used by
Cassisi et al.~(2003) is based on a small set of low mass $\alpha$-enhanced stellar models 
computed with exactly the same code and input physics of the models presented here.
In order to reproduce our initial $Y_{\odot}$ (see previous section) 
we adopt $dY/dZ\sim 1.4$.

In Table~\ref{chemcomp} we summarize the initial chemical composition
of our model grid, and provide also the corresponding [Fe/H] value
(determined assuming the observed solar (Z/X) ratio by Grevesse \& Noels~1993), a
quantity directly comparable to spectroscopic observations. It is
important to notice that the 'solar metallicity' models provide 
[Fe/H]=0.06, instead of 0. The reason is that the model grid does not
include diffusion (which we know is active in the Sun, but, as
discussed before, possibly inhibited at least at the surface of other low
mass stars); when diffusion is included the 'solar metallicity' 
composition would provide
[Fe/H]=0 only at the solar age for solar-like models.
This also means that our model and isochrone grid computed with the solar initial chemical
composition will be unable to match the properties of the Sun, e.g.,
the helioseismologically determined depth of the convective envelope
and the envelope He abundance, the surface $Z/X$ ratio and in general the inner
sound speed profile. This is common to all databases computed without
diffusion, but on the other hand our same solar isochrones are able to
fit  -- as we will show later --  the main sequence 
(and the rest of the observed CMD) of the open cluster M~67, whose stars
display a well established solar photospheric [Fe/H], has an age close
to the solar value, and its reddening and distance are well determined empirically.

\subsection{The evolutionary tracks and isochrones}

\noindent
For each chemical composition, we have computed the evolution of up to
41 different stellar masses. The minimum mass is around $0.50M_\odot$
(slightly varying with metallicity), whilst the maximum one is always equal to $10M_\odot$.
In the near future we will extend the computations to 
both Very Low Mass stars\footnote{This work will be done in
collaboration with A.W. Irvin and D. Vandenberg.} and more massive stars.

In order to increase the accuracy of the numerical interpolations among
different 
evolutionary tracks, needed for computing
isochrones and for population synthesis studies, and to allow a
detailed analysis of specific evolutionary properties, such as the Red
Giant Branch transition (Sweigart, Greggio, \& Renzini 1990), we
adopted a mass step of $0.1M_\odot$ (or lower) for masses below
$2.5M_\odot$, $0.2M_\odot$ for masses in the range
$2.5\le{M/M_\odot}\le3.0$, 
and $0.5M_\odot$ for more massive
models\footnote{The mass grid is slightly more coarse for the model 
grid accounting for core overshooting.}.

Models less massive than $\sim 3M_\odot$ have been computed
starting from the pre-MS phase, whereas the more massive ones
have been computed starting from a chemically homogeneous
configuration on the MS. 
All models, apart from the less massive
ones whose central H-burning evolutionary times are longer than 
the Hubble time, have been evolved until the C ignition or 
after the first few thermal pulses along the Asymptotic Giant Branch
(AGB) phase. We plan in the near future to extend the evolution of low- and 
intermediate-mass stars up to the end of the thermal pulses
along the AGB, using the synthetic AGB technique (see, e.g., Marigo,
Bressan, \& Chiosi~1996 and reference therein).

In case of models undergoing a violent He flash at the RGB tip, we do
not perform a detailed numerical computation
of this phase; instead, we consider the evolutionary values of both the He
core mass ($M_{cHe}$), and the surface He abundance ($He_S$) at the
RGB tip, and accrete the envelope with the evolutionary 
chemical composition until the appropriate total
mass is obtained. After waiting for the thermal relaxation of the
structure, the resulting Zero Age Horizontal Branch (ZAHB) model 
has the same luminosity of the corresponding model evolved through the
He flash, as demonstrated by Vandenberg et al. (2000). 

A subsample of the evolutionary tracks (canonical models) is shown in Fig.~\ref{tracks} 
for all our adopted chemical compositions. It is evident
that all relevant evolutionary phases are properly sampled by the different tracks.

For each chemical composition we have computed additional He burning
models (that is, HB models in addition to the ones obtained with our
chosen mass loss law)
with various different values of the total stellar mass after the RGB phase,
using a fixed core mass and envelope chemical profile, as given by a RGB 
progenitor having an age (at the RGB tip) of $\sim 13$Gyr (see Fig.~\ref{hb}). 
These models allow one to compute the CMD of synthetic HB
populations, where there is a spread in the amount of mass lost along the RGB
phase (like the case of Galactic globular clusters).

The RGB progenitor of these additional HB models has a mass typically equal to
$0.8M_\odot$ at the lowest metallicities, and increasing up 
to 1.0$M_\odot$ for the more metal-rich compositions.
For these additional computations we have adopted a very fine mass spacing, i.e. more than
30 HB models have been computed 
for each chemical composition. A subset of these models in displayed in Fig.~\ref{hb}.
The very fine mass spacing provides a suitable tool for
investigating the pulsational properties of RR Lyrae stars and Pop.II Cepheids
as a function of the cluster metallicity and HB morphology.

All our post-ZAHB 
computations have been extended either to the onset 
of thermal pulses for more massive models, or until the luminosity 
along the cooling sequence of white dwarfs decreases down to
$\log(L/L_\odot)\sim -2.5$, for less massive models.
They allow a detailed investigation of the various evolutionary paths 
following the exhaustion of the central He burning, i.e.: 
i) Extremely Hot Horizontal Branch stars that do not reach the AGB 
and evolve as AGB-manqu\'e stars; ii) Post Early 
AGB stars, which leave the AGB before the onset of thermal pulses, and
iii) {\em bona fide} 
AGB stars, i.e. stars massive enough to experience the AGB thermal pulses.

In order to produce isochrones for a given initial chemical composition, 
the individual evolutionary tracks have been reduced to the
same number of points, according to a well established procedure (see,
e.g., Prather~1976; Bergbusch \& Vandenberg~1992).
Along each evolutionary
track some characteristic homologous points (key points -- KPs) corresponding to well-defined
evolutionary phases have been identified. 
The choice and number of KPs 
had to fulfill two conditions: 1) all main evolutionary
phases have to be properly accounted for; 2) the number of KPs has to be
large enough to allow a suitable sampling of the track morphology, even for 
faster evolutionary phases. The second condition requires also that the number
of points between two consecutive KPs has to be properly chosen.

As it is well known, when passing from stars with a radiative core 
during the central H-burning phase to more massive stars with
convective cores, the track morphology changes
significantly. Therefore the KPs in this phase are
different for stars with radiative or convective cores.
The adopted KPs and the number of points between consecutive KPs
are listed in Table~\ref{keypoints}, whilst in 
Fig.~\ref{figkeypoints} we show the location
of the key points along a typical intermediate mass star and a low-mass He
burning structure.
This choice allows an adequate sampling of the track morphology for the whole evolutionary
database and it guarantees accurate interpolations when using
this database in population synthesis codes.

Since we have exactly the same KPs at all metallicities and
the same number of points distributed between consecutive KPs, all
isochrones at any age and metallicity will have the same number of
points distributed consistently between corresponding evolutionary phases.
This allows one to easily produce isochrones of intermediate metallicities by simply
interpolating on a point-by-point basis among the available isochrones. 
In this way the use of our database in a stellar population synthesis
algorithm is straightforward and simple.
To this purpose, we have performed a numerical test as
follows. Isochrones for various ages and $Z$=0.008 ([Fe/H]$=-$0.35) 
have been computed
by interpolating (on a point by point basis) quadratically 
among the chemical compositions corresponding to
$Z$=0.004 ([Fe/H]=$-$0.66), $Z$=0.01 ([Fe/H]=$-$0.25) 
and $Z$=0.0198 ([Fe/H]=0.06). These isochrones have been
compared to the actual isochrones of the same metallicity ($Z$=0.008,
i.e. [Fe/H]=$-$0.35),
computed from the appropriate evolutionary tracks. We found that a
simple quadratic interpolation scheme is sufficient to reproduce the
evolving masses to within a few thousandth of solar masses, colors
and magnitudes within 0.02~mag or better along most of the
evolutionary phases spanned by the isochrones, log$(L/L_{\odot})$ and 
log$(T_{eff})$ within a few thousandth dex along the isochrones.

The whole set of evolutionary computations (but the additional HB
models discussed above) have been used to 
compute isochrones for a large age range, namely, from
40~Myr up to 12.5~Gyr (older isochrones can also be computed), 
from the Zero Age Main Sequence up to the first thermal pulse on the AGB 
or to the C ignition.
The entire library of evolutionary tracks and isochrones are
available upon request to one of the authors or can be 
retrieved at the URL site http://www.te.astro.it/BASTI/index.php. 
In the near future, we will make available at this URL site a WEB interface
which will allow to any user to compute isochrones, evolutionary tracks
and synthetic Color-Magnitude diagram (see below) for any specified age, stellar mass,
star formation history and so on (Cordier et al. 2004).

Detailed tables
displaying the relevant evolutionary features
for all computed tracks are also available by ftp and at the WEB page
\footnote{Those interested in obtaining more information,
specific evolutionary results as well as additional set of models can
contact directly one of the authors or use the \lq{request form}\rq\ available at the
quoted WEB site.}. We display in 
Tables~\ref{tabevolow} and \ref{tabevoint} a summary of the 
basic information about the [Fe/H]=0.06 canonical models in selected evolutionary
phases; tables like these ones, for all metallicities and
both canonical and overshooting models, are available at the mentioned
WEB address.

The files containing the isochrones provide at each point (2000 points
in total) the initial value of the evolving mass and the actual one
(in principle different, due to the effect of mass loss), 
$\log(L/L_{\odot})$, log$(T_{eff})$,
M$_V$, $(U-B)$, $(B-V)$, $(V-I)$, $(V-R)$, $(V-J)$, $(V-K)$, $(V-L)$, 
$(H-K)$, on the Johnson-Cousins system. The transformation
from $\log(L/L_{\odot})$ and log$(T_{eff})$ to magnitudes and colors 
have been performed adopting a new set of bolometric
corrections and color - $T_{eff}$ transformations specifically computed 
for this work (see the following section for more details).

These isochrones can be used as input data for the fortran code 
$SYNTHETIC~MAN(ager)$, that we have 
written in order to compute synthetic CMDs,
integrated colors and integrated magnitudes of a generic stellar
population with an arbitrarily chosen Star Formation Rate (SFR) and
Age Metallicity Relation (AMR). The program reads a grid of isochrones
for ages between log(t)=7.6 and 10.10 (age t in years), in steps of
0.05 dex, and a file specifying a SFR and AMR. For each time step 
set by the ages of the input isochrones, the program determines from  
the SFR and AMR the number of stars formed at that age, and their metallicity.
Once these quantities are computed, the program selects values for the
masses of the stars formed, according to a prescribed Initial Mass Function (IMF),
and interpolates among the isochrones in the grid 
(linear interpolation in mass, quadratic interpolation in metallicity) in
order to determine the photometric properties for these objects.

The program can also account for a spread in the AMR, photometric errors,
depth effects, reddening, unresolved binaries. 
In Fig.~\ref{figsynth} we show an example of CMD for a synthetic
population produced by
this program. The simulation contains $\sim$90000 stars, for a Star
Formation Hystory typical of the solar neighborhood (Rocha-Pinto et
al.~2000a,b), using a Salpeter initial mass function, 
a 1$\sigma$ photometric error of 0.03 mag and
a 20\% fraction of unresolved binaries (the mass of the secondary
stars in these systems is determined following Woo et al.~2003)
\footnote{In the near future, any user will be enable to run this program 
for generating synthetic CMDs by using a WEB interface at our URL site.}.

\subsection{Color transformations and bolometric corrections}

\noindent
The $UBVRIJHKL$ synthetic photometry was computed with  passbands,
zero-points, and bolometric corrections as described in Castelli (\cite{C98}), 
Castelli (\cite{C99}), and Bessell, Castelli \& Plez (\cite{BCP98}). Updated
grids of ATLAS9 model atmospheres and fluxes (Castelli \& Kurucz \cite{CK02}) were 
used to generate synthetic colors and bolometric corrections
needed to transform from the theoretical to the observational planes.

\subsubsection{The model atmospheres}

\noindent
New grids of ATLAS9 model atmospheres were computed
for solar and several scaled solar metallicities. For this paper we used 
grids with metallicities
[M/H] equal to +0.5, +0.2, 0.0, $-$0.5, $-$1.0, $-$1.5, $-$2.0, and
$-$2.5\footnote{[M/H] denotes the difference in log(Z/H) with respect to
the solar value; for a scaled solar metal distribution [M/H]=[Fe/H].}.
For all the models the microturbulent velocity is $\xi$=2~km~s$^{-1}$. 
The new grids, called ODFNEW, differ from the previous ones quoted as
K95 and C97 
in Castelli (\cite{C99})
in that:

\begin{itemize}
\item{The number of models is 476 in all the grids. They range from 3500~K to
50000~K in \teff\ and from 0.0 to 5.0 in \logg;} 

\item{All the models have the same number of 72 plane parallel 
layers from $\log$$\tau_{Ross}$=$-$6.875 to $+$2.00 at steps of
$\Delta$$\log$$\tau_{Ross}$=0.125;}

\item{All the models are computed with the updated solar abundances from
Grevesse \& Sauval (\cite{GS98})\footnote{This metal distribution is
slightly different from the one adopted for the stellar evolution
computation; however the difference between the two distributions 
is small and does not cause any appreciable inconsistency.} 
and therefore with updated opacity distribution functions (ODFs),
which now include also H$_{2}$O lines;}

\item{All the models are computed with the convection option switched on 
and with the overshooting option switched off. Mixing-length convection with 
l/H$_{P}$=1.25 is assumed for all the models. The convective flux decreases 
with increasing T$_{\rm eff}$ and it naturally becomes negligible for T$_{\rm eff}$
of the order of 9000~K;}

\item{The quasi-molecular absorption H-H$^{+}$ was added to the opacity. 
It affects the ultraviolet flux of metal-poor A-type stars
(Castelli \& Cacciari, 2001).}

\end{itemize}  

\noindent
The largest differences between these models and the previous ones
occur for \teff $\le$ 4500~K mostly owing to the addition of the H$_{2}$O
contribution to the line opacity.

\subsubsection{The colors}

\noindent
The $U$ passband is from Buser (\cite{BU78}), while the $B$ and $V$ passbands
are from Azusienis \& Strai\v zys (\cite{AS69}). The  $R$ and $I$ Cousins colors
have passbands from Bessell (\cite{B90}). The $JKL$ colors were computed 
with passbands from Johnson (\cite{J65}) reported also by Lamla (\cite{L82}).
For the $H$ color the passband from Bessell \& Brett (\cite{BB88}) was 
adopted. 

Zero-points for the $V$ magnitude and for $(U-B), (B-V), (V-I), 
(V-R), (V-J), (V-K), (V-L)$ color indices were set by normalizing the computed
colors for Vega to the observed colors. The Vega model atmosphere
is from Castelli \& Kurucz (\cite{CK94}) (\teff=9550~K, \logg=3.95, [M/H]=$-$0.5,
$\xi$=2.0~km~s$^{-1}$);  the Vega observed V magnitude and  colors
are $V$=0.03, $(U-B)$=0.0, $(B-V)$=0.0 (Johnson et al. \cite{J96}), $(V-I)$=$-$0.005,
$(V-R)$=$-$0.009 (Bessell, 1983),
$(V-J)$=0.0, $(V-K)$=0.0, $(V-L)$=0.0. For $(H-K)$ the computed index
of Sirius was normalized to the observed one $(H-K)$=$-$0.009 (Glass~1997).
The Sirius model is from Kurucz~(1997) (\teff=9850~K, \logg=4.30, [M/H]=$+$0.4,
$\xi$=0.0~km~s$^{-1}$, He/H=+0.5).

After some preliminary comparisons with open cluster MS loci,
we have decided to use only for MS stars with \teff$\leq$4750~K, and only 
for $(B-V)$, $(V-I)$ and $(V-R)$ colors, the empirical color-\teff
relationships provided by Houdashelt, Bell \& Sweigart~(2000).

The bolometric correction BC$_{V}$ is given by: 

$$BC_{V} = -2.5[log(\sigma T_{eff}^{4}/\pi)-log (\int_{\alpha}^
{\beta} S_{V}(\lambda)F(\lambda)d\lambda)]+K$$\\
where S$_{V}$($\lambda)$ is the V passband and F($\lambda$) is the computed flux.
The constant K was defined by normalizing to zero the smallest 
bolometric correction (in absolute value) of the whole synthetic grid computed 
for [M/H]=0.0 and microturbulent velocity $\xi$ =2.0 km~s$^{-1}$. 
It occurs for the model with \teff= 7250~K, \logg=0.5. 
With this normalization, positive values for BC$_{V}$ are avoided. 
The value of BC$_{V\odot}$ is $-$0.203 for a solar model with 
parameters \teff=5777~K, \logg=4.4377, $\xi$=1.0~km~s$^{-1}$. 

By assuming $V_{\odot}$=$-$26.75 (Hayes,1985), the solar absolute magnitude
is M$_{V\odot}$=4.82. Therefore, for the adopted normalization, 
M$_{bol\odot}$ is given by M$_{V\odot}$+BC$_{V\odot}$=4.62.

All the ODFNEW grids of models and colors are available at the URL sites\\ 
http://kurucz.harvard.edu/grids.html and 
http://wwwuser.oat.ts.astro.it/castelli/grids.html.
In Fig.~\ref{figcompcol} we display a comparison of
our 500 Myr (including overshooting) and 10 Gyr [Fe/H]=0.06 isochrones,
transformed to the $M_V-(V-I)$ CMD by using our adopted
transformations discussed before, plus the Yale (Green~1988), NEXTGEN
(Allard et al.~1997) and older ATLAS~9 (Castelli~1999) ones.  
This comparison is representative of the general differences in the
various color planes covered by our isochrones.
The Yale transformations tend to produce a bluer lower MS, RGB and
He-burning phase, whereas the cooler and brighter part of the RGB tends
to agree with our results. Color differences can be as high as
$\sim$0.2 mag. The NEXTGEN transformations are available for surface
gravities higher or equal to log(g)=3.5 (in cgs units), and are
different (bluer) only on the lower MS.
The earlier ATLAS~9 transformations are very similar apart from the
lower MS and cooler RGB, where color differences of $\sim$0.2 mag or
more are reached.
\section{Comparison with existing databases}

\noindent
In this section we compare the H-R diagram of selected isochrones with the
counterparts from the publicly available databases by Girardi et al.~(2000), Castellani
et al.~(2003), Lejeune \& Schaerer~(2001) and Yi et al.~(2001). 
In addition, we add a comparison with Bergsbusch \& Vandenberg~(2001) 
isochrones kindly provided by the authors (D.A. VandenBerg~2003, 
private communication). 

These various grids of
models are computed employing some different choices of the physical
inputs (opacities are usually the same, but EOS, nuclear reaction
rates, boundary conditions and neutrino energy loss rates are often different), 
and also the chemical compositions are generally different.
We perform the comparison on the theoretical H-R diagram, thus bypassing
the additional degree of freedom introduced by the choice of the
colour transformations. We employed our isochrones including
overshooting (unless otherwise stated), 
since all these sets of isochrones are computed
including overshooting from the convective cores, albeit often with
different prescriptions about its extension and the decrease to zero 
for decreasing stellar mass.

We have selected isochrone
ages that span all the relevant age range, and 
chemical compositions as close as possible to our choices. It is
important to notice that Castellani et al.~(2003) and Yi et al.~(2001)
models do include atomic
diffusion (He and metals for Castellani et al. models, only He for
Yi et al. models), and therefore their isochrones for old ages (above a few Gyr) are
affected by the efficiency of this process.

Figure~\ref{compPadua} shows a comparison with the Girardi et
al.~(2000) isochrones; the chemical composition of our isochrones is
as labeled, very close to the $Z$=0.019, $Y$=0.273 composition of
Girardi et al.~(2000) isochrones. The selected ages are 100~Myr, 500~Myr, 
1.8~Gyr and 10~Gyr, respectively. The MS loci do agree reasonably
well, whereas our RGBs are systematically hotter, apart from the
youngest isochrone. Our He-burning models are slightly brighter,
especially for the 500~Myr isochrone, but become fainter at 100~Myr.
The TO regions are in good agreement at 10 Gyr; our TO
models become brighter at 1.8 Gyr, have the same luminosity as
Girardi et al.~(2000) but are hotter at 500~Myr, and 
are fainter at 100~Myr. At 1.8~Gyr the TO region of Girardi et al.~(2000) 
isochrones is reproduced if we select an age higher by 0.7~Gyr (e.g., the mass
evolving at the TO of our isochrone is higher than the Girardi et
al.~2000 counterpart); at 100~Myr we need
to choose an age 10~Myr younger to reproduce the TO of Girardi et
al.~(2000) isochrones.
Since Girardi et al.~(2000) computed also a set of canonical
isochrones for this metallicity, we have investigated in more detail
the reason for the relatively large difference at 1.8 Gyr. We
discovered that the difference between the 1.8 Gyr old canonical
isochrones --  which are only due to the difference in the model input physics
-- accounts for about 40\% of the discrepancy between the overshooting
models. The remaining fraction of the discrepancy has to be ascribed
to the interplay between the overshooting treatment and the different
input physics.

In Fig.~\ref{compPisa} we display a comparison with selected 
isochrones without overshooting and initial $Z$=0.008 and $Y$=0.250, 
from the database by Castellani et al.~(2003). The agreement with our
canonical models is quite good all along the MS, TO region and
RGB. Our He-burning luminosities are slightly lower, and at 10 Gyr 
there is a larger difference at the TO region due to the effect of
atomic diffusion included by Castellani et al.~(2003).
We do not find a substantial discrepancy in the TO
brightness at 1.8~Gyr, as in case of the comparison with the canonical
models by Girardi et al.~(2000).

The comparison with Lejeune \& Schaerer~(2001) models with 
initial $Z$=0.004 and $Y$=0.252 is displayed in 
Fig.~\ref{compGeneva}. Ages are the same as in the previous comparison.
The isochrone MSs of the two sets look in reasonable agreement, and the TO region of
the oldest and youngest isochrones are practically identical. However
for the 500~Myr and 1.8~Gyr the TO of our isochrones is clearly
brighter. At 1.8 Gyr we should employ an isochrone about 1 Gyr older in
order to match the  Lejeune \& Schaerer~(2001) one.
Our RGBs are much hotter and the He-burning phase (when available in
Lejeune \& Schaerer~2001 isochrones) is brighter at 500~Myr and
fainter at 100~Myr. 

In case of the Yi et al.~(2001) database, we compared with the 
$Z$=0.02, $Y$=0.27 isochrones, for ages of 100~Myr, 600~Myr, 1.8~Gyr
and 10 Gyr (see Fig.~\ref{compY2}). Also in this case the MS loci appear to be in good
agreement, apart from the lower luminosities, where Yi et al.~(2001)
models show a different slope. The TO of their 10 Gyr isochrone is
affected by diffusion, and behaves accordingly when compared with our
models. At 100 and 500~Myr the brightness of the TO regions is
similar, although effective temperatures are slightly different; 
at 1.8 Gyr there is again a difference in the TO luminosity, that translates into
an age difference of about 800~Myr, as in case of the comparison with
Girardi et al.~(2000) isochrones; our RGBs are hotter at all ages.

The comparison with VandenBerg isochrones  ($Y$=0.277 and $Z$=0.0188) is shown in 
Fig.~\ref{compVdB}, for ages of 2 and 10~Gyr. The 10~Gyr isochrones do
agree very well also along the RGB. At 2~Gyr we find the same
differences encountered in the other comparisons.

A common result of the previous discussion is the fact that our RGB
temperatures are generally hotter than the others, and the TO
luminosity for our 1.8 Gyr isochrones is higher.
As far as the RGB temperatures are concerned, the difference is certainly not due 
(apart from the comparison with Vandenberg models, where the RGB
location is exactly the same in the 10 Gyr isochrones which have the 
same RGB evolving mass, and with Castellani et al. models)
to different value of the mass evolving along the RGB (higher masses when our
isochrones are brighter at a given age), since we find this
difference also with respect to isochrones with the same RGB evolving mass.
We will see in the next section that this difference in the RGB
temperatures -- possibly due to a combination of different EOS, and
solar mixing length calibration obtained using different boundary
conditions, see e.g. Salaris, Cassisi \& Weiss~(2002) -- 
does not prevent our models from fitting very well the
observed RGBs in a sample of Galactic open clusters, when coupled to 
our color transformations. 
To justify better why we think these effective temperature
discrepancies along the RGB may be due to a combination of different
boundary conditions and EOS, we notice that 
in Salaris, Cassisi \& Weiss~(2002) it has been shown how
different boundary conditions (for example grey vs Krishna Swamy
$T(\tau)$ relationship) produce different solar mixing length calibrations that
make all solar models agree at the Sun location, but disagree along the RGB 
by differences of the order of 100 K. As for the influence of
the EOS, the same 1$M_{\odot}$ solar composition models discussed in
\S2, computed using the Irwin EOS and the EOS employed in Cassisi \&
Salaris~(1997) keeping everything else unchanged, show $T_{eff}$
differences along the RGB of the order of $\sim$ 280 K at the base of
the RGB, $\sim$140 K at log($L/L_{\odot}$)=2 and $\sim$85 K at the He flash
ignition, in the sense of having hotter temperatures with the Irwin EOS.

As for the TO brightness differences
at 1.8 Gyr (which however is negligible in case of the comparison
with Castellani et al.~2003 models without overshooting), 
it persists for ages higher than 1.8
Gyr, up to when the overshooting distance and the size of the
convective cores are reduced to zero (ages of about 5--6 Gyr). 
We believe that this might be related 
mainly to the different MS lifetime caused by the use of a different
EOS, coupled to differences in the way the overshooting distance is
reduced to zero when approaching the radiative core regime. 

To conclude this section, we show in Fig.~\ref{comphb} a comparison
between our ZAHB luminosity at the instability strip vs [Fe/H]
theoretical relationship, and analogous results from various authors. 
Our new values are systematically brighter than Cassisi \&
Salaris~(1997) models, due mainly to the effect of the new plasma neutrino
rates and the higher initial He content (in Cassisi \& Salaris~1997 we
employed a primordial He mass fraction equal to 0.23). 
The main difference in Cassisi et al.~(1998) models is the EOS for the electron
degenerate cores along the RGB, that increases the He flash core mass with
respect to our result (they also used the I93 electron
conduction opacities
that slightly increase the He core mass in this metallicity range),
but this increase is offset by the fact that their adopted He 
abundance is lower.
Vandenberg et al.~(2000) models are the faintest ones of this group, and we
believe this is mainly due to their use of the Hubbard \& Lampe~(1969)
electron conduction opacities (that produce a lower He core mass at the He
flash with respect to our adopted opacities), 
coupled to a lower primordial He abundance.
The results by Caloi et al.~(1997) have an overall different slope
with respect to the other models, and we are not able to exactly
pinpoint the reason for this difference.


\section{Empirical tests}

\noindent
In this section we present the results of some empirical test we
performed in order to establish the consistency of our models with 
photometric constraints. We have successfully tested our models in
Cassisi et al~(2003) and Riello et al.~(2003), in the regime of
Galactic globular clusters. In particular, we have shown that with our
models the initial He abundance of globular clusters as estimated 
from the $R$-parameter, is finally in agreement with the cosmological
constraint (Cassisi et al.~2003). More tests on Galactic globular cluster
stars (whose initial chemical composition is characterized by
[$\alpha$/Fe]$>$0) will be presented in the companion paper with the
$\alpha$-enhanced model grid. 
The reason for this is twofold. First, 
scaled solar isochrones are not an appropriate surrogate for
$\alpha$-enhanced ones with the same $Z$ when $Z$ is above
$\approx$0.002 (e.g. Salaris \& Weiss~1998); second, from our 
new $\alpha$-enhanced color transformations we have found that. e.g.,  
scaled solar $(B-V)$-effective temperature relationships are different from
$\alpha$-enhanced ones, starting from relatively low metallicities around
$Z\sim$ 0.001.

In the comparisons detailed below, we employed the CMDs of field stars
and a sample of clusters with empirically established 
parameters like distance, [Fe/H] (and scaled solar metal mixture) and reddening.
In this way we do not have much freedom to change the
parameters of our models in order to match the observations. Comparisons
with, e.g., LMC clusters would be much less meaningful, since
reddenings, [Fe/H] (especially) and distances (not only the distance to
the LMC barycenter, but also geometrical effects) are not well
established empirically. We could fit the CMDs of those clusters and
give our estimates of their distance and reddening by playing with
combinations of reddening, distance and [Fe/H], but this would not be a real test for the 
accuracy of the models.

We also refrained from comparing theoretical integrated colors with 
galactic open clusters or LMC data. 
The problem is that integrated colors of galactic open
clusters are very uncertain due to their low total mass that induces 
large stochastic color fluctuations preventing any meaningful
comparisons (see, e.g., the numerical experiments by Bruzual \& 
Charlot~2003 and the integrated colors reported in the
WEBDA database -- Mermilliod~1992 -- at http://obswww.unige.ch/webda/integre.html).
The same applies to LMC clusters, with the additional problems of not
well known age and uncertain reddening. The situation is better
for the more massive Galactic globular clusters, and in fact we
will perform (using the appropriate $\alpha$-enhanced theoretical
isochrones and color transformations) this comparison in our forthcoming 
paper about $\alpha$-enhanced models.

In the following comparisons, 
when necessary, we will use the
reddening law $A_V=3.24E(B-V)$, $A_I=1.96E(B-V)$, and
$A_K=0.11A_V$ (Schlegel, Finkbeiner \& Davis~1998).

\subsection{Solar neighborhood MS stars}   

Percival et al.~(2002) have published homogeneous multicolor photometry and 
photometric metallicity estimates for a large sample of local field dwarfs with 
accurate $Hipparcos$ parallaxes; the [Fe/H] values of the individual
stars range from $\sim-$0.4 up to +0.3. 
These data are extraordinarily well suited to test the
MS location of our theoretical isochrones in various colour bands; moreover, 
the choice of the isochrone age is irrelevant, since the observed stars
are all unevolved, i.e. fainter than $M_V \sim$5.5. 

We have divided the Percival et al.~(2002) sample into three
subsamples, as follows. The most metal poor subsample comprises
objects with [Fe/H] between $-0.35$ and $-$0.15, with an average value
of [Fe/H]=$-0.24\pm0.06$; the intermediate subsample contains stars
with [Fe/H] between $-$0.05 and +0.15, with an average value of
0.06$\pm$0.06. The third subsample is made of the most metal rich
objects, with [Fe/H] between 0.16 and 0.36, and an average [Fe/H]
equal to 0.26$\pm$0.09.
In Figs.~\ref{subdwbv} to \ref{subdwjk} we compare the CMD of these three subgroups with
our isochrones for [Fe/H]=$-0.25$, 0.06 and +0.26, respectively. The
comparison is made in the $M_V-(B-V)$, $M_V-(V-I)$, $M_V-(V-R)$ and 
$M_K-(V-K)$, $M_K-(J-K)$ planes; the $J$ and $K$ magnitudes of the field stars come
from the $2MASS$ database (M. Groenewegen \& S. Percival, private
communication), and have been transformed to our adopted IR system 
following Carpenter~(2001). For the sake of comparison we also
show a Girardi et al~(2000) isochrone with [Fe/H]=0.06.
 
The agreement between the empirical CMDs and the isochrones appears to
be satisfactory -- 
in the sense that the theoretical isochrones usually overlap the main
CMD locus occupied by the dwarfs in the various metallicity bins
and color planes-- 
although it tends to deteriorate slightly -- apart from the case of 
$M_K-(V-K)$ -- at the highest metallicity 
(where, however, the star sample is the smallest).

\subsection{Open clusters}

We have employed a sample of galactic open clusters of various ages 
for further tests of the reliability of the magnitudes and colors 
of our isochrones. More in detail, we have considered 
the CMDs of the Hyades (Pinsonneault et al.~2003), 
NGC~2420 (Anthony-Twarog et al.~1990), M~67 (Sandquist~2004) 
and NGC~188~($BV$ from Caputo et al.~1990, and $VI$ from  
von Hippel \& Sarajedini~1998), for which precise parallax-based
distance determinations do exist. In case of the Hyades accurate 
parallaxes of the individual stars in the observed CMD have been
determined by the $Hipparcos$ satellite. For the other clusters the
distances have been determined by Percival \& Salaris~(2003 -- PS03)
using a completely empirical MS-fitting method presented in Percival
et al.~(2002). The MS template sequences used in this technique have
been computed making use of the field dwarf sample by Percival
et al.~(2002), whose individual [Fe/H] values are on the same scale of
the open cluster metallicities by Gratton~(2000).
PS03 employed clusters' [Fe/H] estimates from Gratton~(2000) and reddening from 
Twarog et al.~(1997).

When we compare our models to the observed CMD of a cluster, we have chosen
within our isochrones grid the one with metallicity closest to the
cluster [Fe/H]. 
However, if no computed isochrone has a [Fe/H] value within the quoted
1$\sigma$ error associated to the cluster one, we interpolate quadratically within our grid 
as previously discussed, in order to determine the isochrone with the 
precise cluster metallicity.

We will consider a fit 'successful' when the theoretical
isochrones overlap the main cluster CMD branches 
for values of reddening, metallicity and distance modulus in agreement with the
empirical determinations.

Figures~\ref{Hyadesbvi} and \ref{Hyadesvk} display the comparison 
with the Hyades CMD corrected for the cluster distance; 
the cluster metallicity is [Fe/H]=0.13
according to Gratton~(2000), and the reddening is zero.
The observed CMD contains exactly the same stars in all three
different photometric planes.
The interpolated isochrone with [Fe/H]=0.13 
that best matches the Hyades CMD has an age 
of 790~Myr if overshooting is included, or 560~Myr for canonical
models; 
notice the match (that in $(B-V)$ and $(V-K)$ 
degrades only when $M_V >$ 7.5)
of the unevolved MS to the observed one, without applying any colour shift.
The TO region in $K-(V-K)$ is however less well reproduced with 
respect to the $BVI$ CMD, a fact that can
be ascribed either to some inconsistencies in the color transformations,
or photometric data or a combination of both.

In case of NGC~2420, PS03 obtained $(m-M)_0=11.94\pm0.07$, 
assuming [Fe/H]=$-$0.44 and E$(B-V)=0.05\pm$0.02; our interpolated
isochrones with [Fe/H]=$-$0.44 that best match the observed CMD provide
a distance modulus $(m-M)_0=11.90$ and E$(B-V)$=0.06, in good
agreement with the MS-fitting results (see Fig.~\ref{2420bv}). 
The age of the best fitting
isochron with overshooting is 3.2~Gyr, whereas the canonical one has an
age of 2~Gyr. 
It is interesting to notice that the isochrone with
overshooting does provide a better match of the observed CMD TO region.

As for M~67, PS03 have determined a distance modulus 
$(m-M)_0$=9.60$\pm$0.09, adopting E$(B-V)$=0.04$\pm$0.02 and 
[Fe/H]=0.02$\pm$0.06. Their MS-fitting determination was based on a
different $BVI$ photometry (Montgomery et al.~1993), 
however Sandquist~(2004) has obtained
exactly the same distance modulus with his photometry, employing the
same method by PS03 and the same [Fe/H] and E$(B-V)$.
We fitted our [Fe/H]=0.06 isochrones to the $BVI$ cluster data, and obtained a
successful fit 
with E$(B-V)$=0.02 and $(m-M)_0$=9.66, in agreement, within the error bar,
with the MS-fitting results. The result of the fit is shown 
in Fig.~\ref{67bvi}. The MS and RGB 
of the cluster appear to be 
matched by both
overshooting and canonical isochrones; also the brightness of the
He-burning stars is reproduced. 
It is also worth noticing that Sandquist~(2004, his Fig.~14) has shown how a number
of isochrones in the literature is unable to fit the main location of
the MS for $V > 15.5-16$, when the empirical distance, reddening and
metallicity are considered; our isochrones, however, do provide a much
better match to this MS region.
The ages of the best fitting
isochrones are 4.8 Gyr and 3.8 Gyr for the overshooting and canonical
models, respectively.
As already noticed by Sandquist~(2004) who employed the Yi et al.~(2001)
and the Girardi et al.~(2000) isochrones, the overshooting models do
provide a worse fit to the TO region compared to the canonical
ones. The mass evolving at the TO of the overshooting isochrone 
is equal to $\sim$1.2~$M_{\odot}$,
e.g., it is in the mass interval where the amount of overshooting is
decreasing with mass. This means that M~67 seem to indicate that
the extension of the overshooting region has to be already almost down
to zero for masses of $\sim$1.2~$M_{\odot}$, at least for
metallicities around solar.
Figure~\ref{67vjk} shows a comparison between the same isochrones displayed
in Fig.~\ref{67bvi} and M~67 CMDs in near-infrared colors. 
Subgiants and RGB stars appear to be consistent with our
isochrones for the same age, reddening and distance modulus necessary 
to fit the $BVI$ data, with a small mismatch in the $K$ band
absolute magnitude of He burning stars. 

PS03 have determined for NGC~188 $(m-M)_0=11.17\pm$0.08,
by using  E$(B-V)=0.09\pm$0.02 and [Fe/H]=$-0.03\pm$0.06.
We fitted our interpolated [Fe/H]=$-0.03$ isochrones to the observed CMD, 
in both $M_V-(B-V)$ and $M_V-(V-I)$; the
successful fit to 
the whole observed CMD is achieved for an age of 6.3~Gyr
(for both isochrones with and without overshooting), E$(B-V)$=0.09 
and a distance modulus $(m-M)_0$=11.17, coincident with the
empirical MS-fitting result. 
Figure~\ref{188bvi} shows the result of the fit with theoretical isochrones;
the main location of the MS 
is reproduced and one can also 
easily notice that the isochrone with overshooting is only slightly different
from the canonical one. This is due to the fact that the value of the 
mass evolving at the
TO region is of about 1.16~$M_{\odot}$, and for these masses
the extension of the
overshooting region is reducing to zero in our models. Thus, the
comparison with NGC~188 does not contradict the possibility, raised by
the comparison with M~67, that the overshooting extension is already
reduced to zero when the stellar mass is $\sim1.2-1.3M_{\odot}$ for
metallicities around solar.

We have also compared our isochrones with the deep CMD by
Kalirai et al.~(2001) of NGC~6819, a cluster not employed by PS03
(see Fig.~\ref{6819bv}). We have adopted
the recent [Fe/H]=0.09$\pm$0.03 determination by Bragaglia et
al.~(2001), and found a 
a successful fit with our
[Fe/H]=0.06 isochrones for E$(B-V)$=0.14 -- in perfect agreement 
with the spectroscopic value E$(B-V)$=0.14$\pm$0.04 
determined by Bragaglia et al.~(2001) -- and a distance modulus 
$(m-M)_0$=12.10. The age of the cluster is 3 Gyr when overshooting is
accounted for, or 1.8 Gyr in case of canonical isochrones.
The isochrones overlap the main locus of the cluster MS, and also the brightness of
the He-burning clump stars is matched.
In this case the isochrone with overshooting does provide a
better match to the TO region.

\subsection{The eclipsing binary AI~Phoenicis}

The eclipsing binary star AI~Phoenicis is a particularly well suited
candidate to test the calibration of the mixing length adopted in our
computations, and the amount of overshooting from the convective
core. The eclipsing binary analysis provides extremely precise values
for the masses (1.190$\pm$0.006 $M_{\odot}$ and 1.231$\pm$0.005 $M_{\odot}$) and 
radii (1.762$\pm$0.007 $R_{\odot}$ and 2.931$\pm$0.007 $R_{\odot}$) 
of the two components (Milone, Stagg \& Kurucz~1992); 
moreover, the metallicity of the system is estimated
spectroscopically to be [Fe/H]=$-0.14\pm$0.10 by Andersen et al.~(1988).
The less massive component is located right after the TO region, at
the beginning of the subgiant branch, whereas the more massive star is
in a more advanced phase, close to the base of the RGB; this means that
their radii are very sensitive to both the mixing length calibration
and overshooting treatment.
In Fig.~\ref{AIPhe} we display a comparison on the mass--radius plane
between the observed values for the system, and the best match
theoretical isochron computed for [Fe/H]=$-$0.14 (quadratically 
interpolated among the available grid of metallicities). Both components can
be fitted within the extremely small 1$\sigma$ error bars with the
same age (approximately the solar age) and without including
overshooting. Isochrones with overshooting cannot match simultaneously
both components (an age of 5.37~Gyr provides the closest possible
match to both components, that is however much worse than the case of
canonical isochrones), even though, with our prescription, the overshooting
region is of about 0.08$H_P$ in this mass range.
This result agrees with the case of M~67, where we have shown
(in agreement with Sandquist~(2004) results obtained by employing different 
isochrones) that masses of $\sim 1.2M_{\odot}$ have an overshooting region
already reduced to virtually zero for metallicities around solar.

\section{Summary}

\noindent
We have presented an homogeneous and updated database of stellar evolution
models and isochrones for old-, intermediate-, and young stellar
populations for a wide range of chemical compositions.
Two large grids, with and without the inclusion of
convective core overshooting during the H-burning phase are provided. 
This set of theoretical models is suitable
for population synthesis analysis as well as for investigating evolutionary
properties of field and cluster stars in a wide range of masses and chemical compositions.
The main points of this investigation can be summarized as follows:

\begin{itemize}

\item{our models have been computed making use of many of the most
recent updates of the stellar physics inputs.
In particular, we employed new improved electron conduction opacities that 
cover the relevant physical conditions of electron degenerate cores
along the RGB; new homogeneous reaction rates from the NACRE
compilation, plus the latest determination for the important 
$^{12}$C$(\alpha,\gamma)^{16}$O reaction; 
new improved ATLAS9 color-transformations and bolometric corrections;
new EOS that covers homogeneously all the stellar structure along all the relevant
evolutionary phases for the entire mass range covered by our
computations. This EOS is in extremely good 
agreement with the OPAL EOS in the common domain of validity,  
and eliminates the need to match the OPAL EOS to a different one when 
dealing with He burning and more advanced evolutionary stages.
}

\item{the adopted initial He abundance for metal-poor models is in agreement
with the most recent evaluations in this field obtained by CMB analysis: this value
is also consistent with the estimate obtained from the R-parameter
analysis on a large sample of Galactic globular clusters, performed 
with a theoretical calibration consistent with present models. 
This consistency is not achieved by the other available databases.
}

\item{The results of our calculations allow the computation 
of grids of isochrones covering a wide range of stellar ages, 
as well as the estimate of several astrophysical parameters such as the 
RGB transition mass, the He core mass at the He ignition, the amount of 
extra-helium dredged up during  RGB evolution, the properties
of models at the beginning of the thermal pulse phase on the AGB or
at the carbon ignition;}

\item{we have computed also an additional large set of HB
models -- starting from a He core mass typical of ages around 13
Gyr -- for all the metallicities available in our grid, that can be used to
produce synthetic HB CMDs of various morphologies;}

\item{stellar models and isochrones have been transferred to
several observational CMDs by adopting updated bolometric corrections 
and color - $T_{eff}$ relations, computed specifically for this
project, supplemented by empirical color transformation for the lower
MS in selected color indices. The evolutionary tracks and isochrones
are normalized in such a way that a simple interpolation on a point by
point basis can be used in population synthesis codes to produce
isochrones for metallicities intermediate among the values provided in the grid;}

\item{we have shown  various comparisons with empirical data for
unevolved field MS stars with
available $Hipparcos$ precise parallax measurements and metallicity
estimates, and with data for open clusters with accurate
photometry, metallicity estimates and empirical MS-fitting
distances. As a whole, one can notice that a good
agreement has been achieved between theory and observations;} 

\item{we have carefully treated the way in which the
core overshoot efficiency is decreased
when approaching the radiative core regime in low-mass stars.
Our analysis of the open cluster M~67 and the eclipsing binary AI Phoenicis 
shows that, at least around solar metallicities, core overshooting efficiency should
be reduced to zero already for masses of the order of
$1.2-1.3M{_\odot}$. Further empirical tests should be
performed in order to reach a definitive conclusion on this issue. We
suggest always to compare results obtained with both canonical and
overshooting models and isochrones for ages of the order of 4-5 Gyr.}

\end{itemize}

\noindent
We have made a large effort in order to make available the whole theoretical framework
to the scientific community in a easy and direct way through a dedicated WEB site. We
plan, in the near future, to prepare a WEB
interactive interface that will allow any potential user to compute
isochrones, luminosity functions, synthetic CMDs and integrated
colors for any specified age and/or star formation history. The evolutionary
database will be soon implemented with a self-consistent and extended set of theoretical
models for an $\alpha-$enhanced heavy element mixture, that will be presented in a
forthcoming paper. 

\begin{figure}        
\plotone{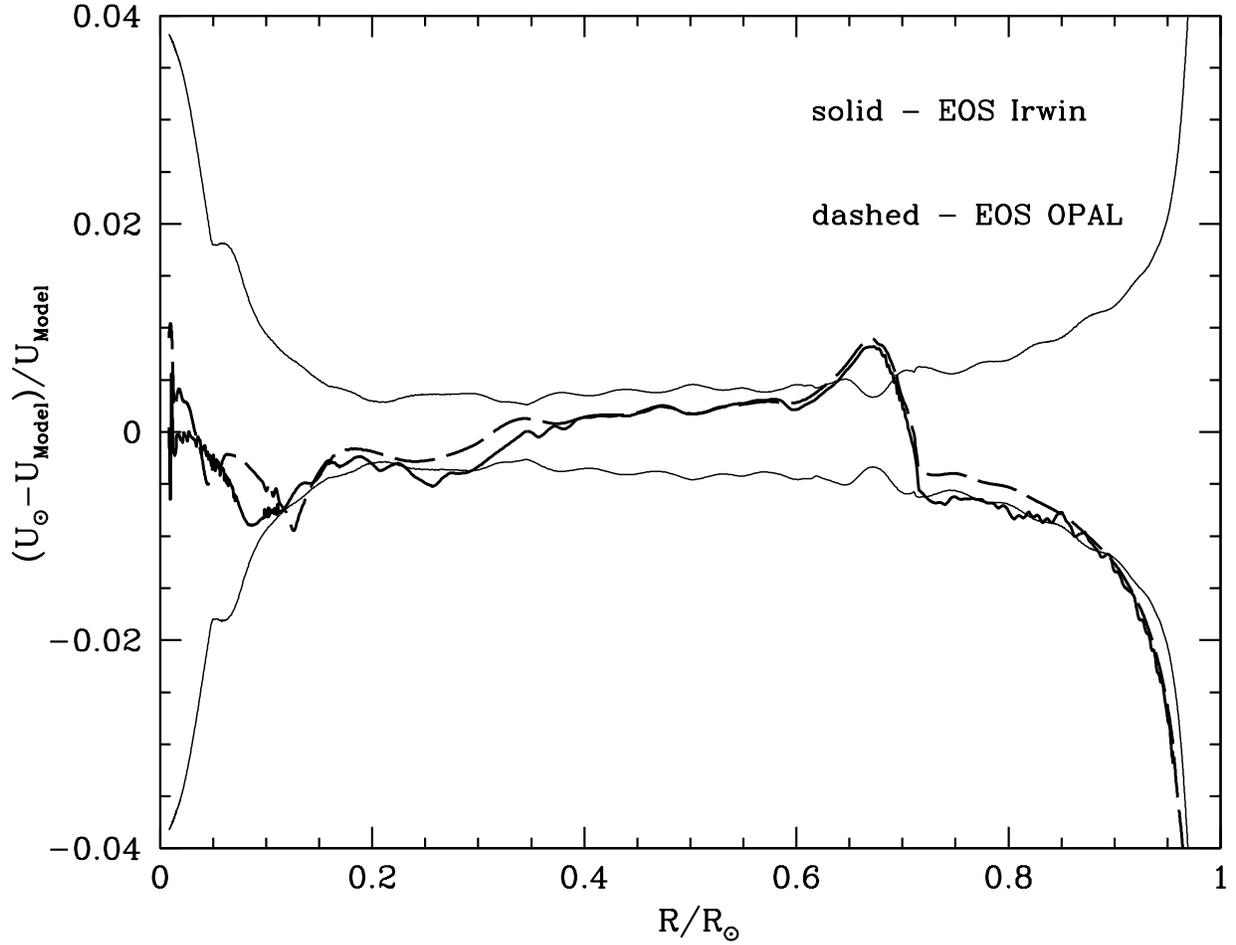}        
\caption{Relative differences between the helioseismologically
inferred solar U=P/$\rho$ profile (Degl'Innocenti et al.~1997) 
and the results from two
theoretical solar models computed using the OPAL EOS and the Irwin
EOS, respectively. Thin
solid lines mark the ``conservative'' error on the helioseismological
result as discussed by Degl'Innocenti et al.~(1997).
\label{sspeed}}        
\end{figure}
\begin{figure}        
\plotone{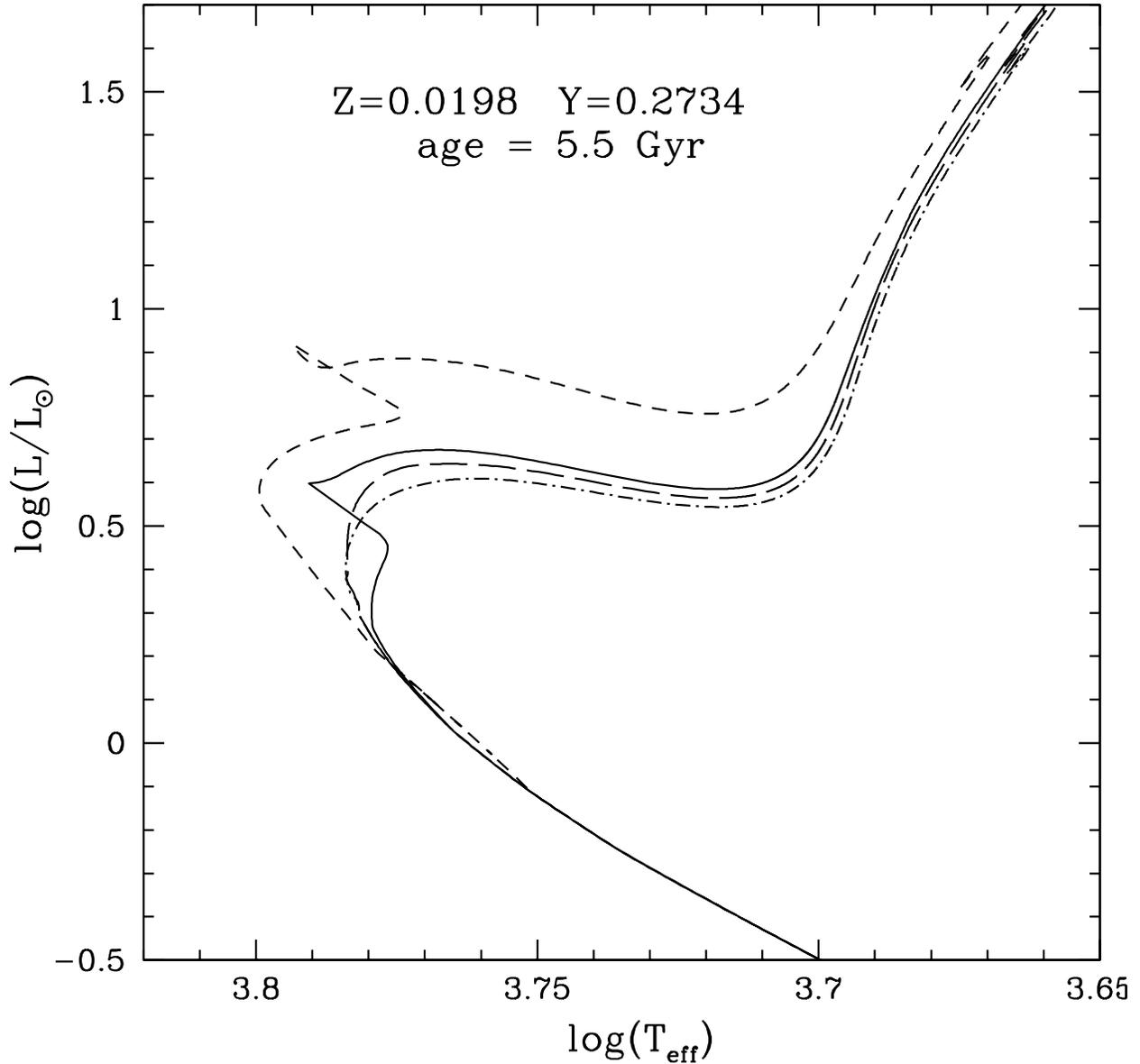}        
\caption{Theoretical isochrones with the same (see labels) age and
chemical composition, computed with different ways to decrease 
the core overshoot efficiency with stellar mass: the dot-dashed line
corresponds to canonical models; the solid one has been obtained from 
models computed using the prescription adopted in this work (see text for more details);
the short-dashed one
has been obtained by assuming that $\lambda_{OV}$ is equal to 0.20$H_P$ for stellar 
masses larger
than 1.3$M_\odot$, equal to 0 for $M\le0.95M_\odot$, and linearly
decreasing in the intermediate
range of masses; the long-dashed one has been obtained by assuming 
that $\lambda_{OV}$ is equal to 0.20$H_P$ for stellar masses larger
than 1.7$M_\odot$, equal to 0.15$H_P$ for $1.5M_\odot$, vanishing for $1.1M_\odot$ and 
linearly decreasing in the intermediate range.
\label{over0}}        
\end{figure}
\begin{figure}        
\plotone{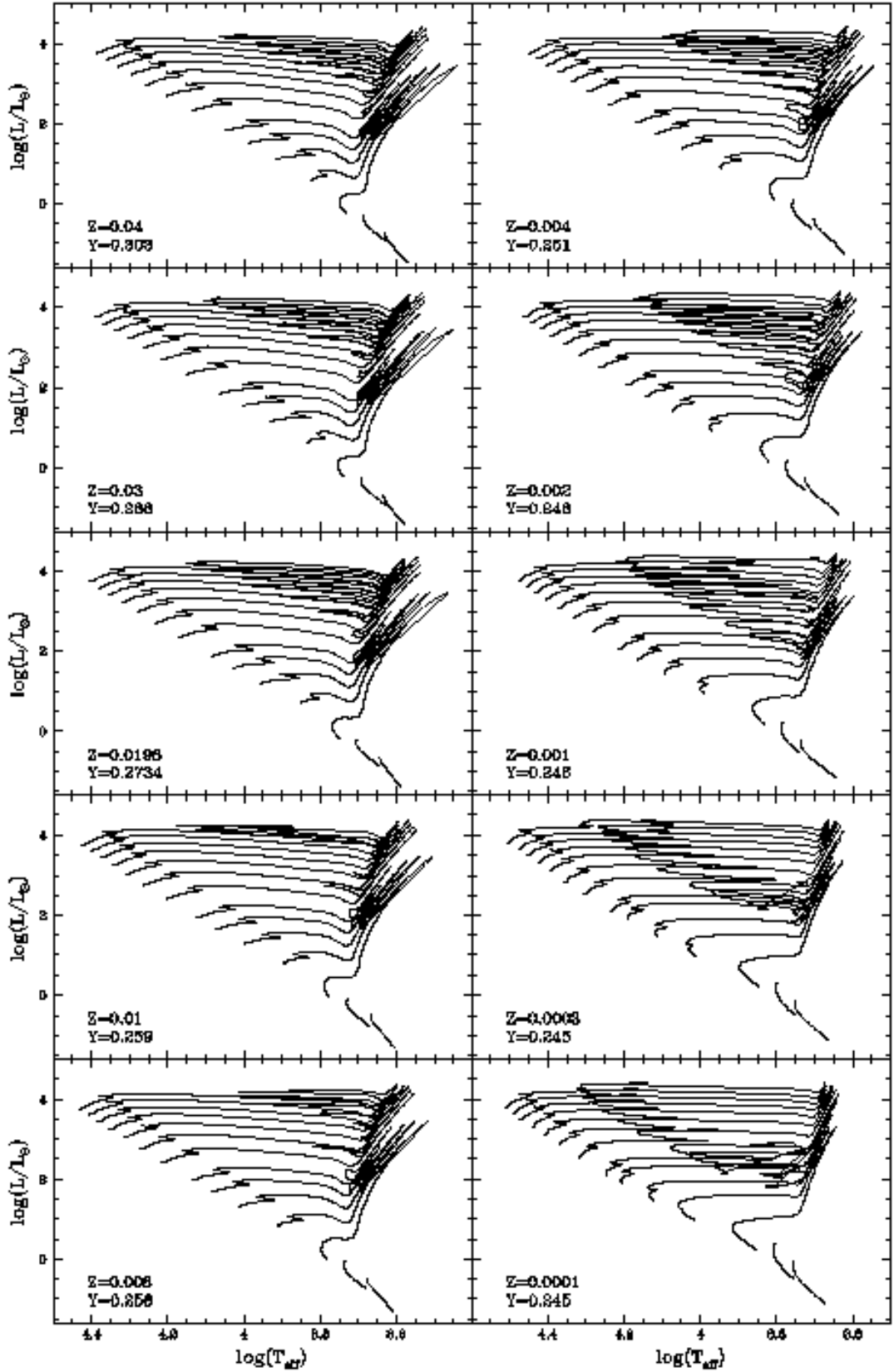}        
\caption{Selected evolutionary tracks for the 10 metallicities of our
model grid.  The plotted tracks correspond to models with the following initial
masses: $M/M_\odot=$ 0.5, 0.7, 1.0, 1.5, 2.0, 2.5, 3.0, 4.0, 5.0, 6.0, 7.0, 8.0, 9.0 and
10.0. 
\label{tracks}}        
\end{figure}
\begin{figure}        
\plotone{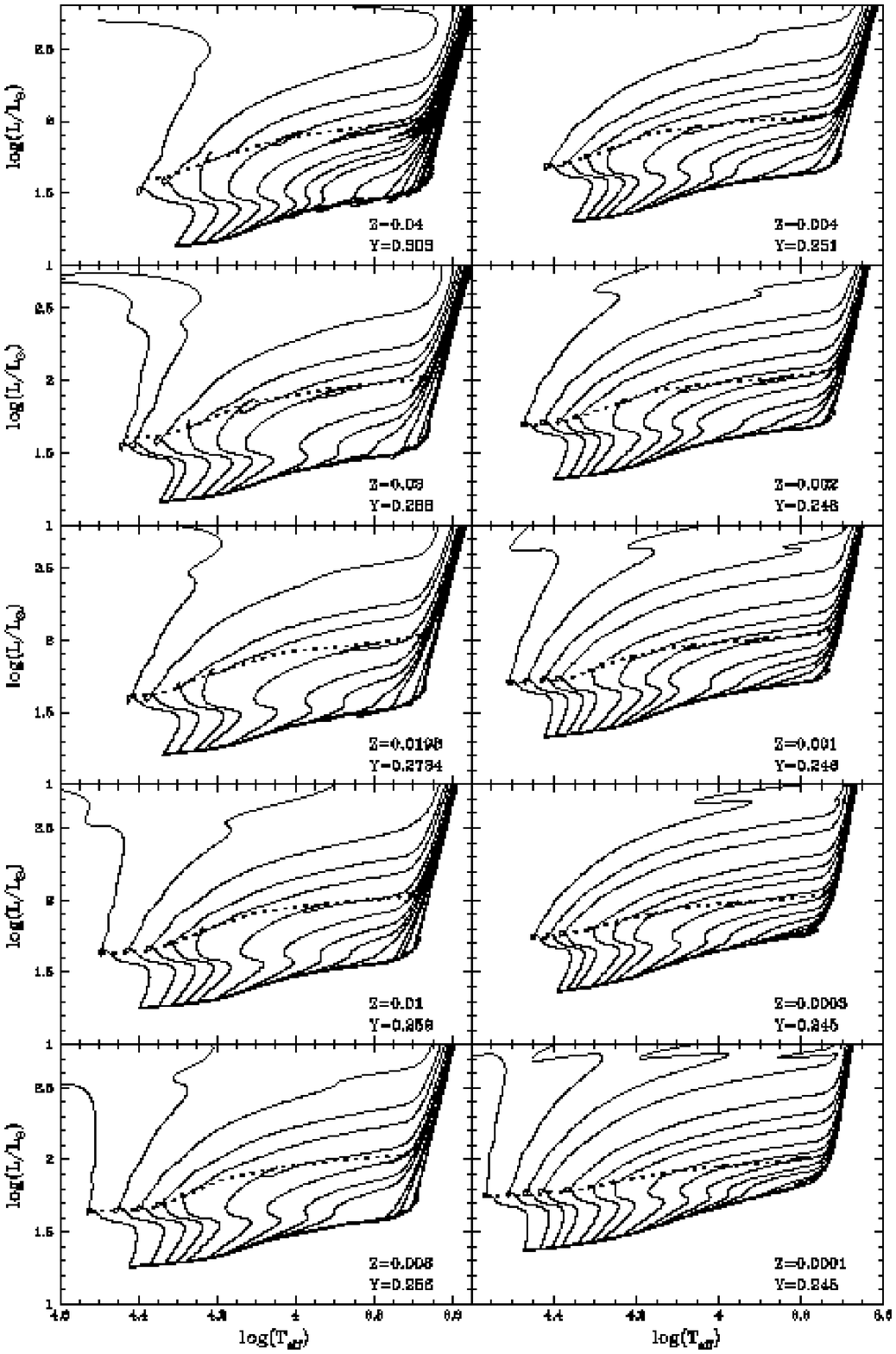}        
\caption{Additional HB evolutionary tracks suited for the computation
of synthetic HB populations. The figure shows also the location of the ZAHB and central He
exhaustion loci.
\label{hb}}        
\end{figure}
\begin{figure}        
\plotone{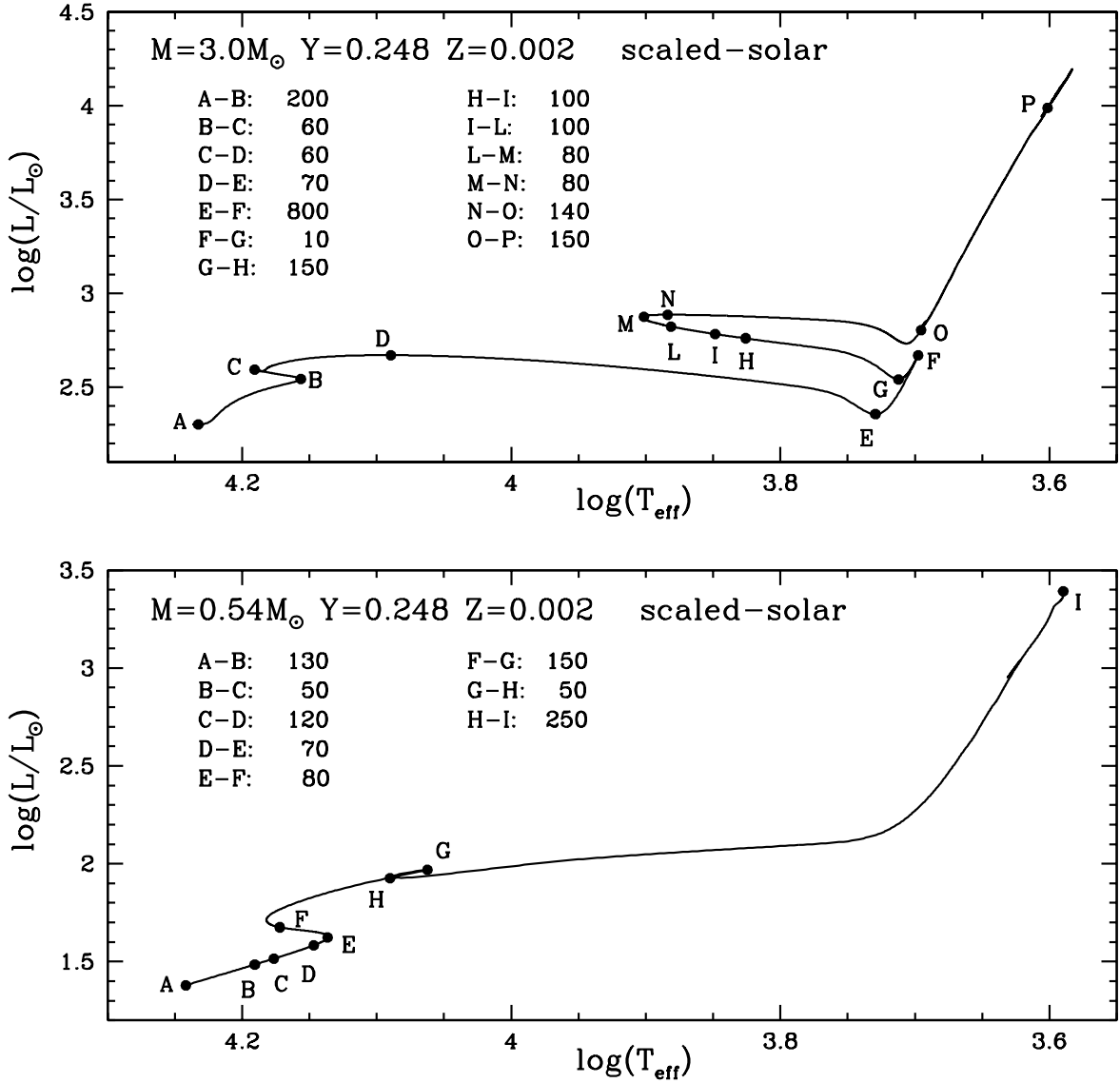}        
\caption{Location of the Key Points along selected evolutionary tracks.
\label{figkeypoints}}        
\end{figure}
\begin{figure}        
\plotone{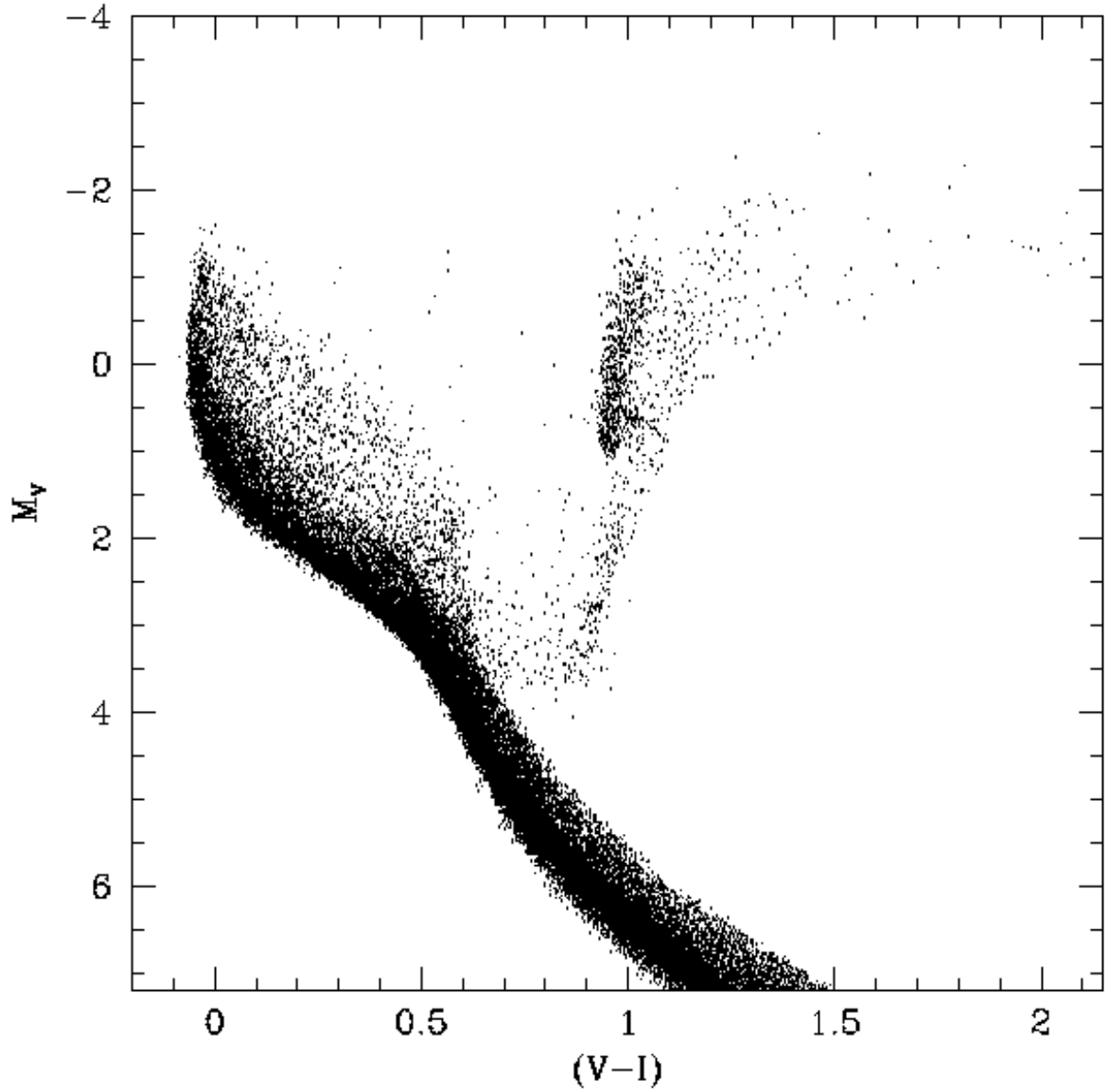}        
\caption{Example of CMD produced by 
our code SYNTHETIC~MAN for a composite stellar population (see text for details).
\label{figsynth}}        
\end{figure}
\begin{figure}        
\plotone{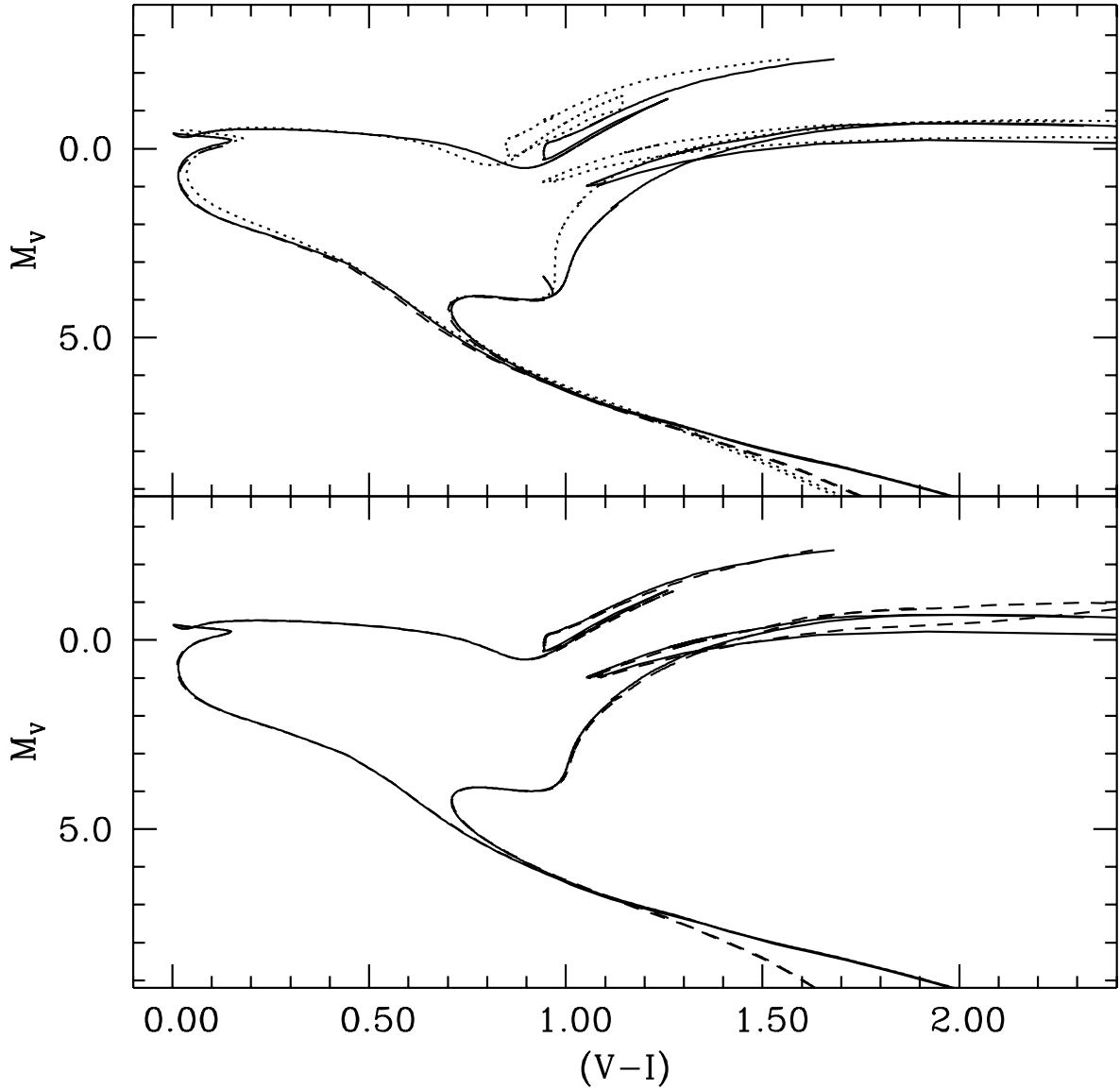}        
\caption{Comparison between our [Fe/H]0.06 
isochrones of 500 Myr (including
overshooting) and 10 Gyr transformed to the $M_V-(V-I)$ CMD by 
using various sets of
color-transformations, i.e., the transformations adopted in this paper
(solid line), the Yale transformations by Green~(1988 -- upper panel
dotted line), NEXTGEN transformations by Allard et
al.~(1997 -- upper panel dashed line), and 
the earlier ATLAS~9 transformations by Castelli~(1999 -- lower panel 
dashed line).
\label{figcompcol}}        
\end{figure}
\begin{figure}        
\plotone{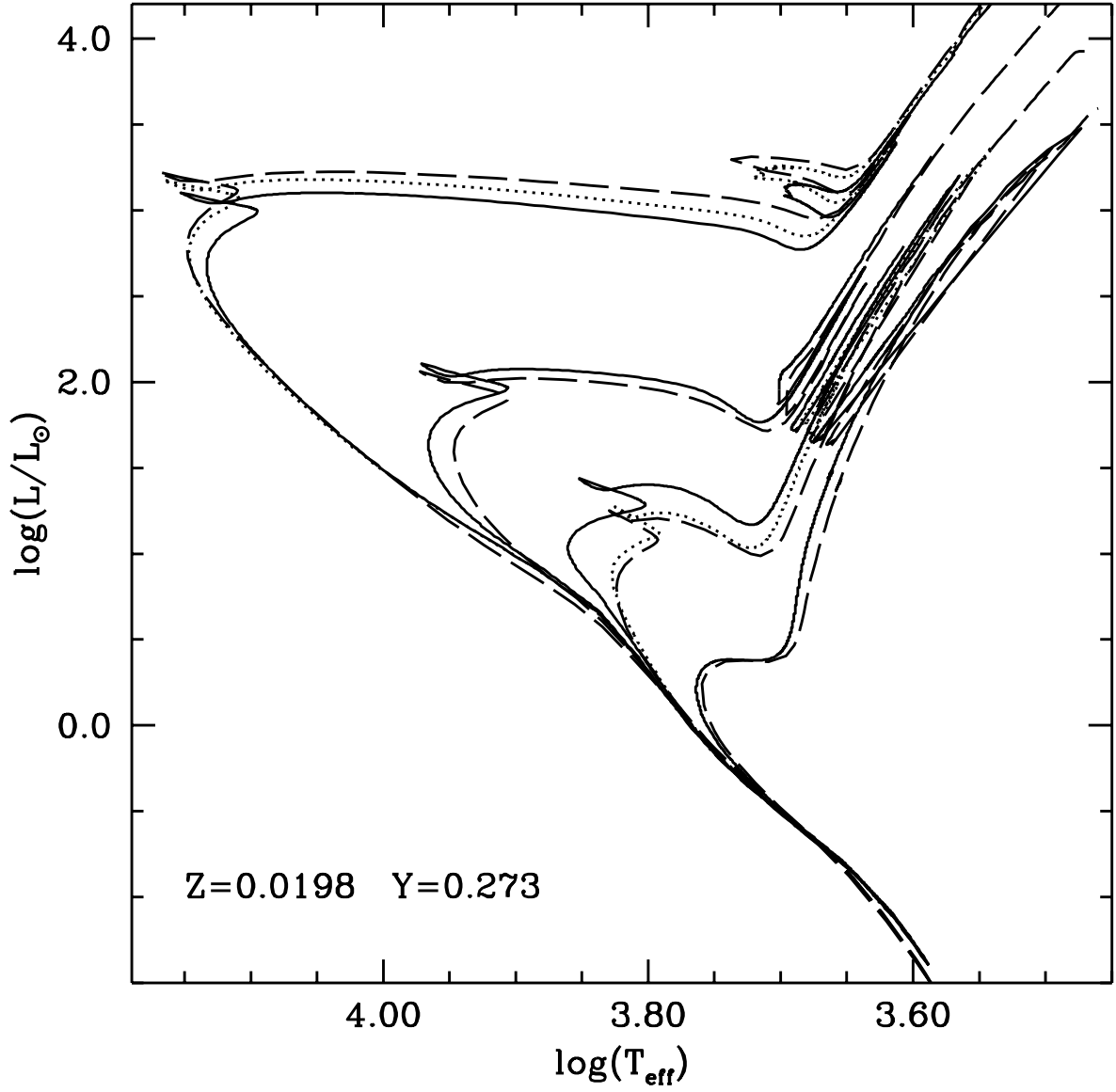}        
\caption{Comparison between our isochrones (solid lines) and Girardi et al.~(2000)
isochrones (dashed lines) for selected ages (100~Myr, 500~Myr,
1.8~Gyr, 10~Gyr). The chemical composition of our
isochrones is as labeled; Girardi et al.~(2000) isochrones have
$Y$=0.273 and $Z$=0.019. Dotted lines show our isochrones for ages of  
90 Myr and 2.5 Gyr, respectively.
\label{compPadua}}        
\end{figure}
\begin{figure}        
\plotone{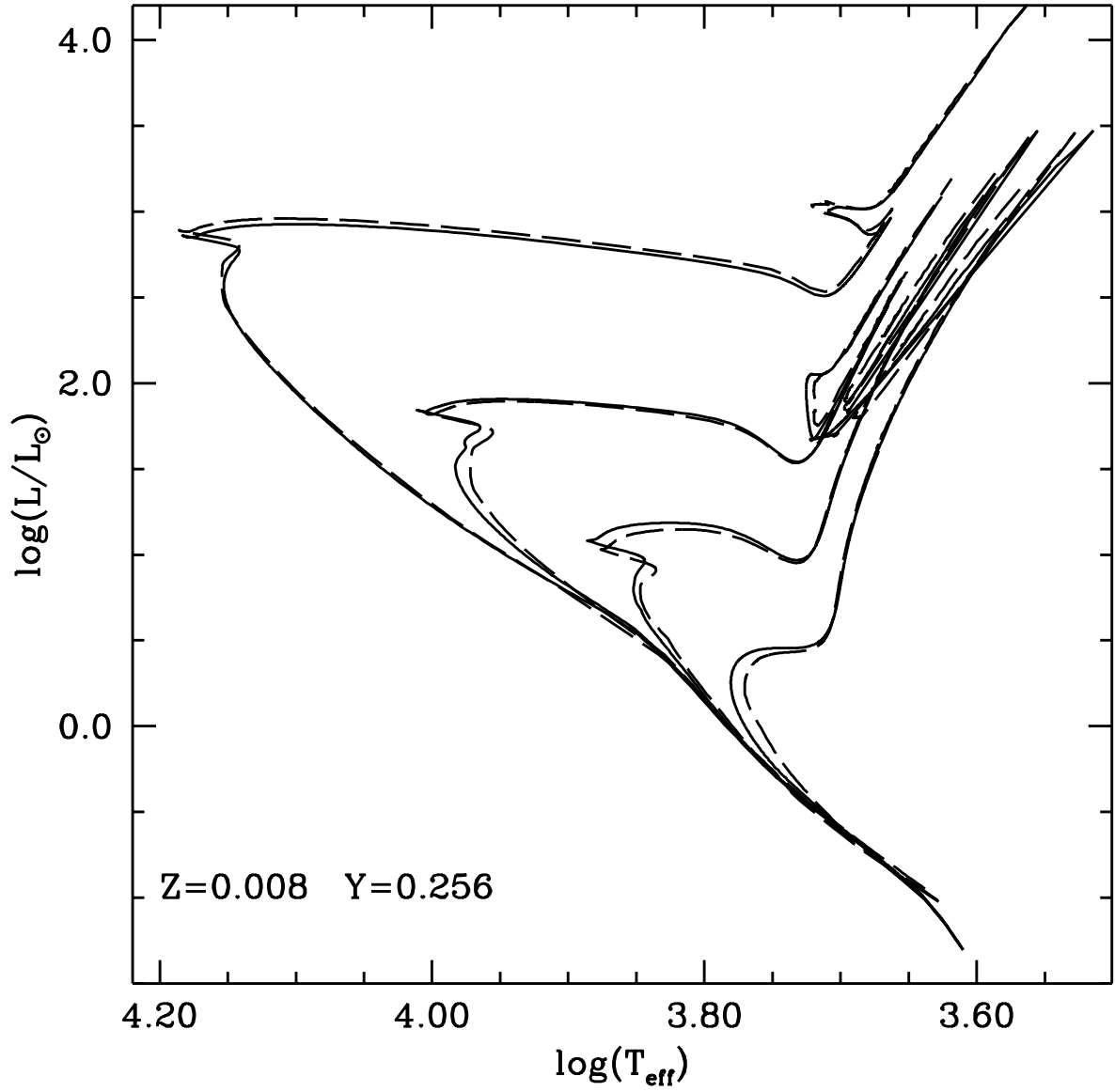}        
\caption{As in Fig.~\ref{compPadua} but for the isochrones by
Castellani et al.~(2003) with $Y$=0.250 and and $Z$=0.008.
\label{compPisa}}        
\end{figure}
\begin{figure}        
\plotone{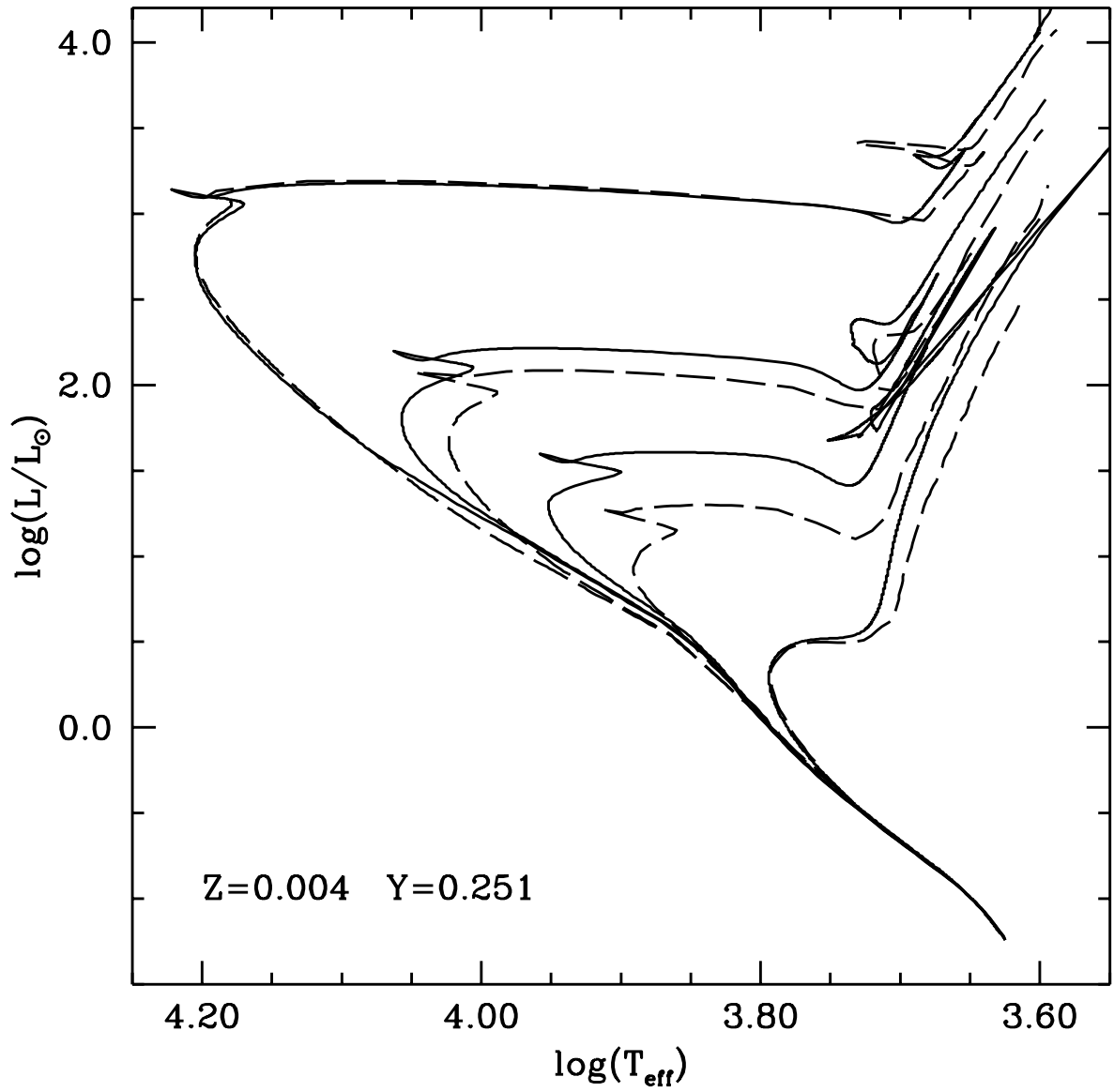}        
\caption{As in Fig.~\ref{compPadua} but for the isochrones by Lejeune
\& Schaerer~(2001) with $Y$=0.252 and and $Z$=0.004.
\label{compGeneva}}        
\end{figure}
\begin{figure}        
\plotone{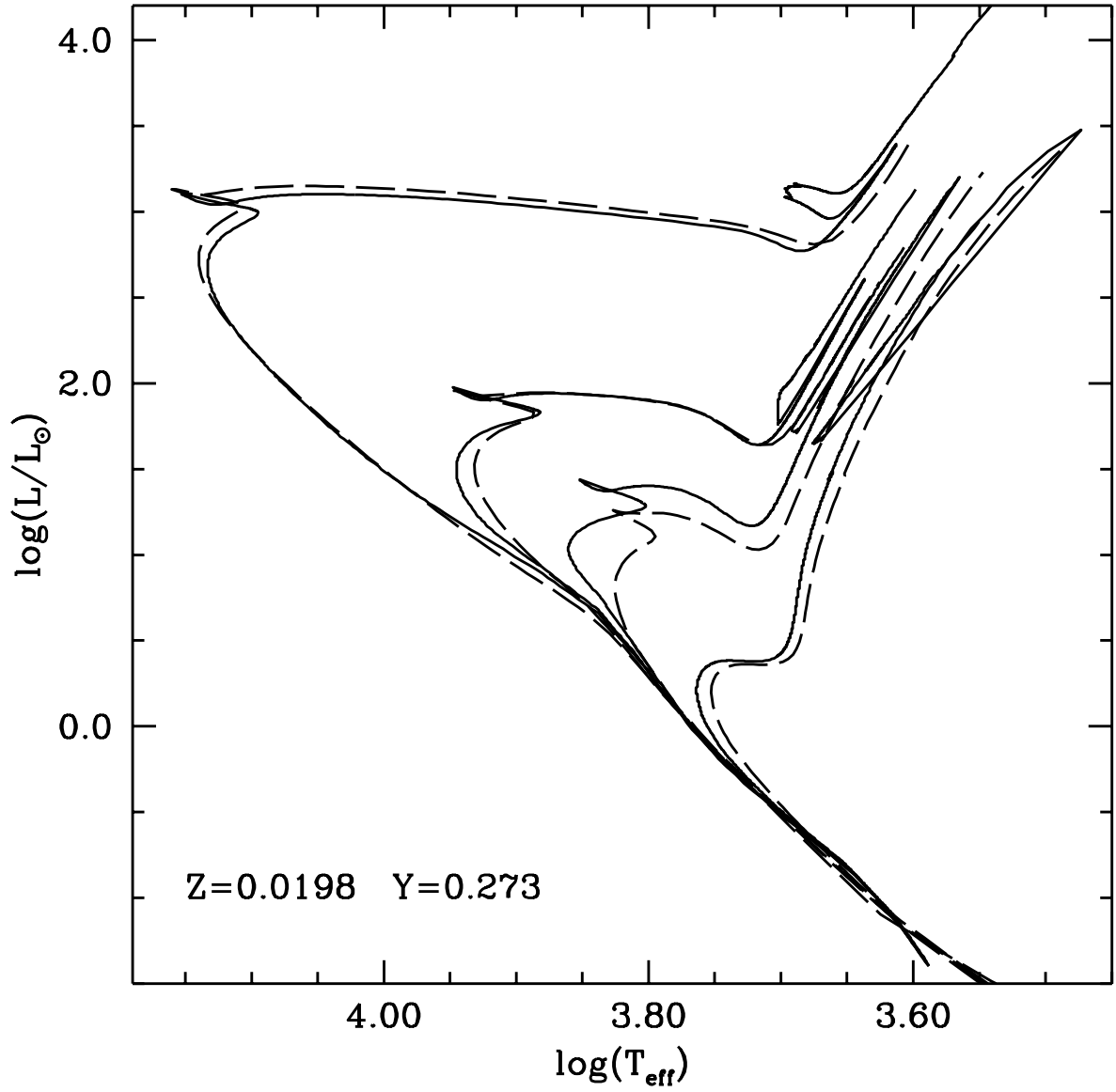}        
\caption{As in Fig.~\ref{compPadua} but for the isochrones by Yi et
al.~(2001) with $Y$=0.27 and and $Z$=0.02. Selected ages are 100~Myr,
600~Myr, 1.8~Gyr and 10~Gyr, respectively.
\label{compY2}}        
\end{figure}
\begin{figure}        
\plotone{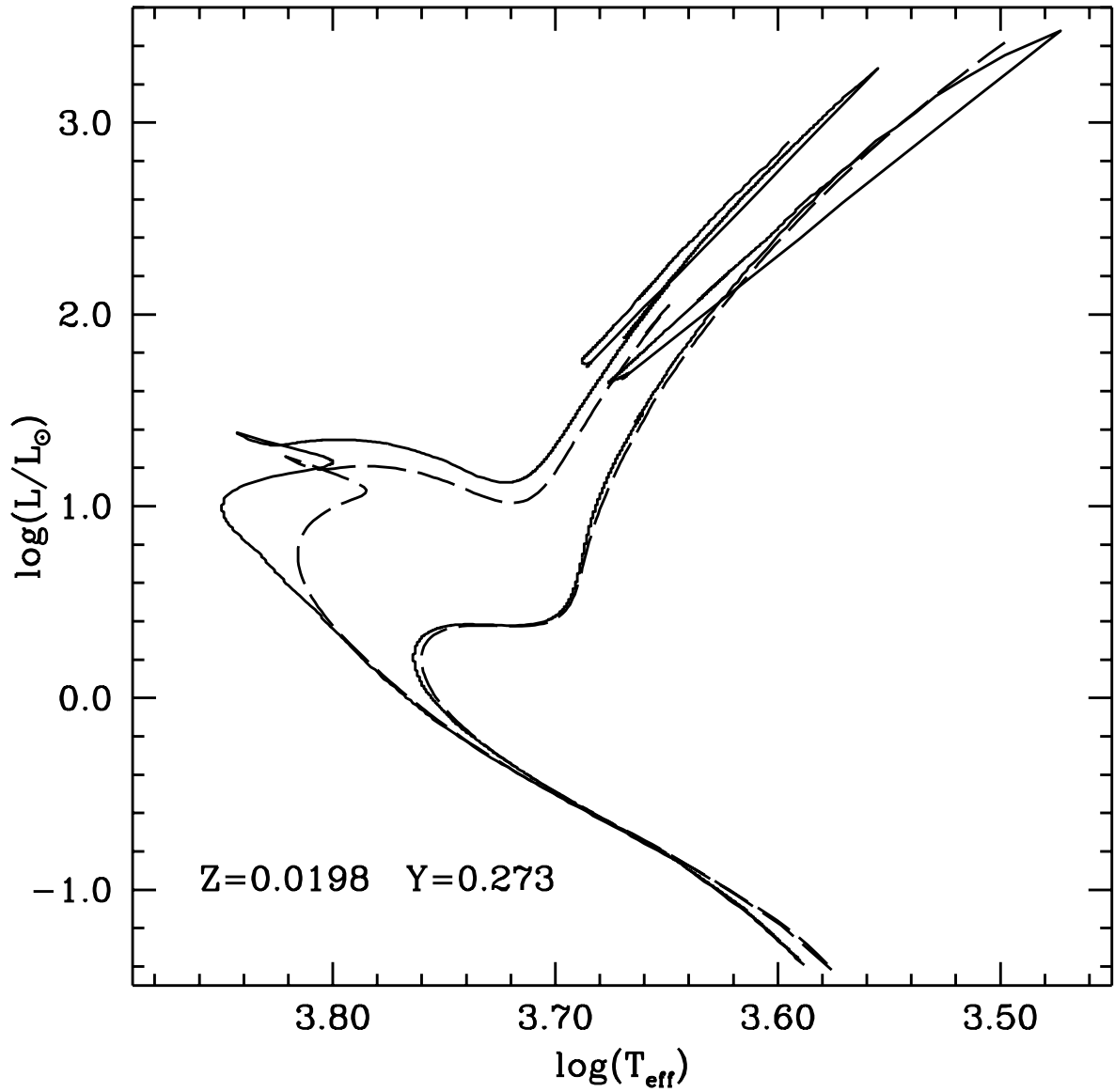}        
\caption{As in Fig.~\ref{compPadua} but for VandenBerg (private
communication) isochrones with $Y$=0.277 and $Z$=0.0188. Selected ages
are 2~Gyr and 10~Gyr, respectively.
\label{compVdB}}        
\end{figure}
\begin{figure}        
\plotone{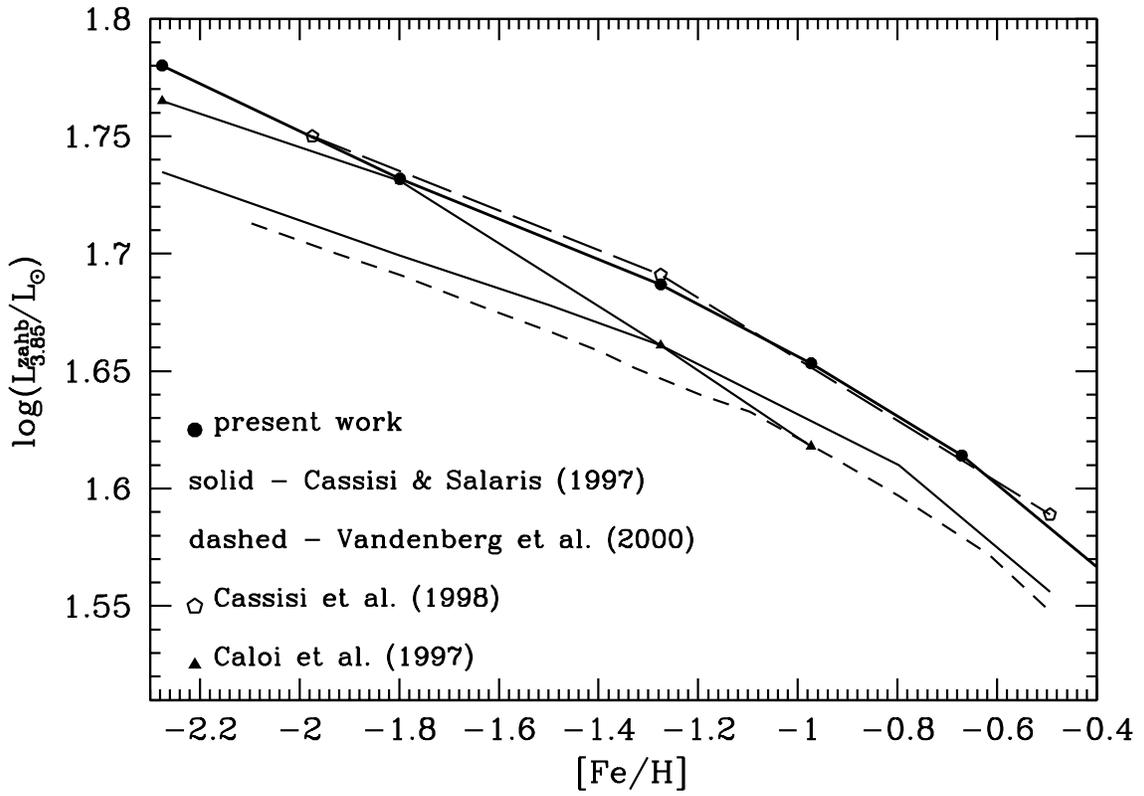}        
\caption{Comparison of the labeled theoretical relationships between
the ZAHB luminosity at the instability strip (log($T_{eff}$=3.85) and [Fe/H].
\label{comphb}}        
\end{figure}
\begin{figure}        
\plotone{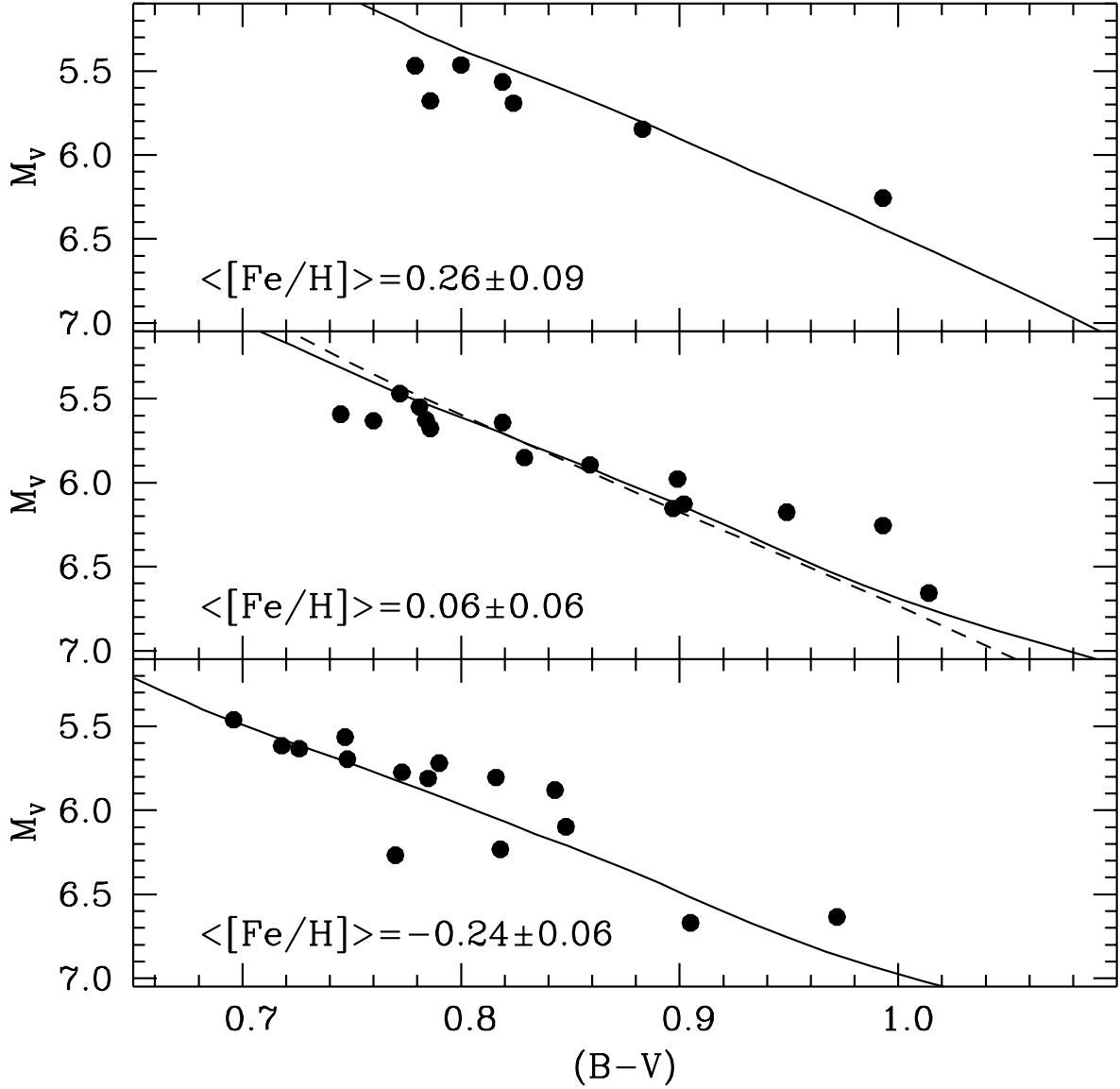}        
\caption{Comparison in the $M_V-(B-V)$ CMD 
between the unevolved MS of our isochrones and
field dwarfs with the labeled mean metallicities (see text for details). The dashed line
refers to Girardi et al. (2000) isochrone with [Fe/H]=0.06.
\label{subdwbv}}        
\end{figure}
\begin{figure}        
\plotone{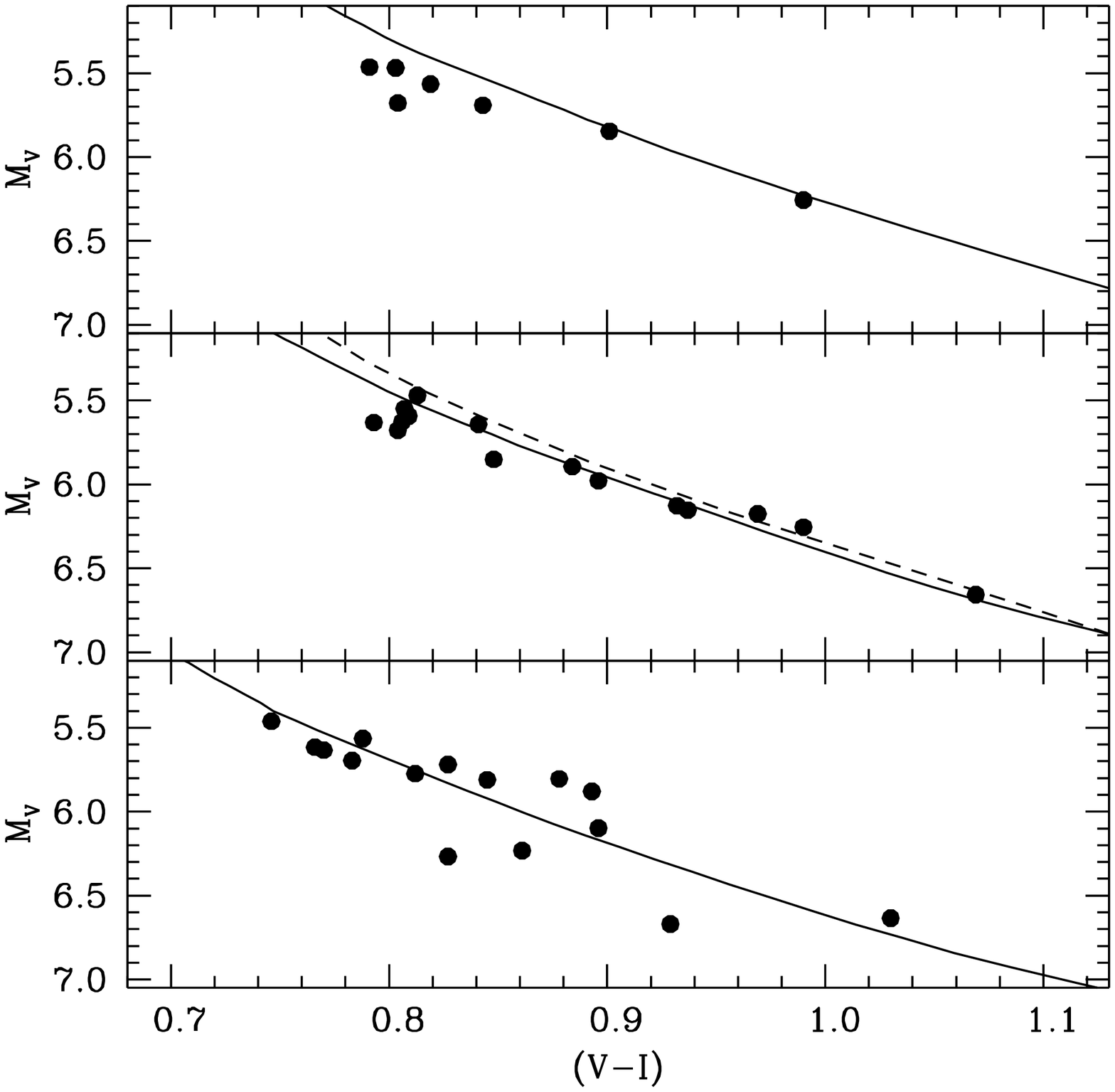}        
\caption{As in Fig.~\ref{subdwbv} but for the $M_V-(V-I)$ CMD.
\label{subdwvi}}        
\end{figure}
\begin{figure}        
\plotone{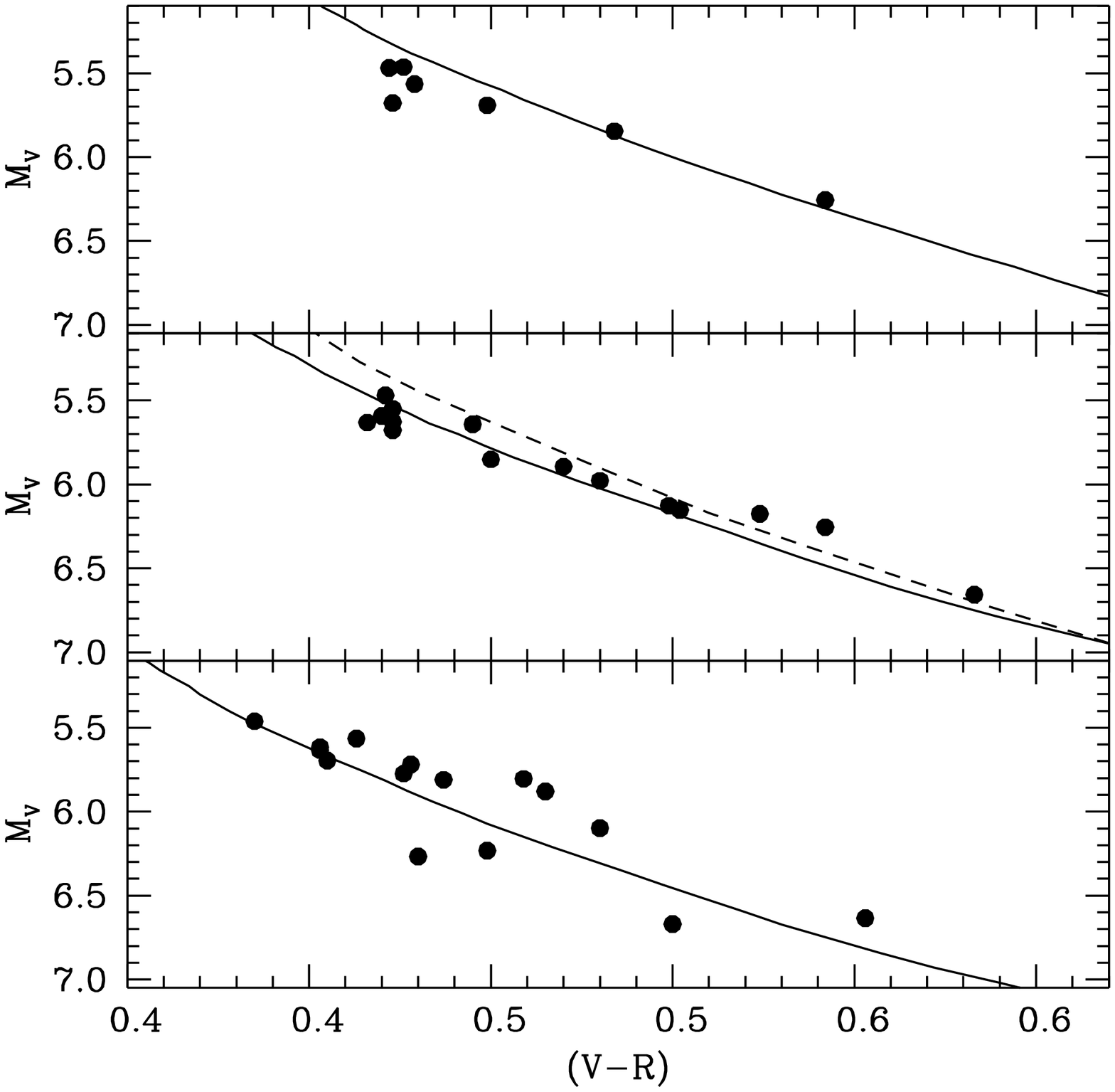}        
\caption{As in Fig.~\ref{subdwbv} but for the $M_V-(V-R)$ CMD.
\label{subdwvr}}        
\end{figure}
\begin{figure}        
\plotone{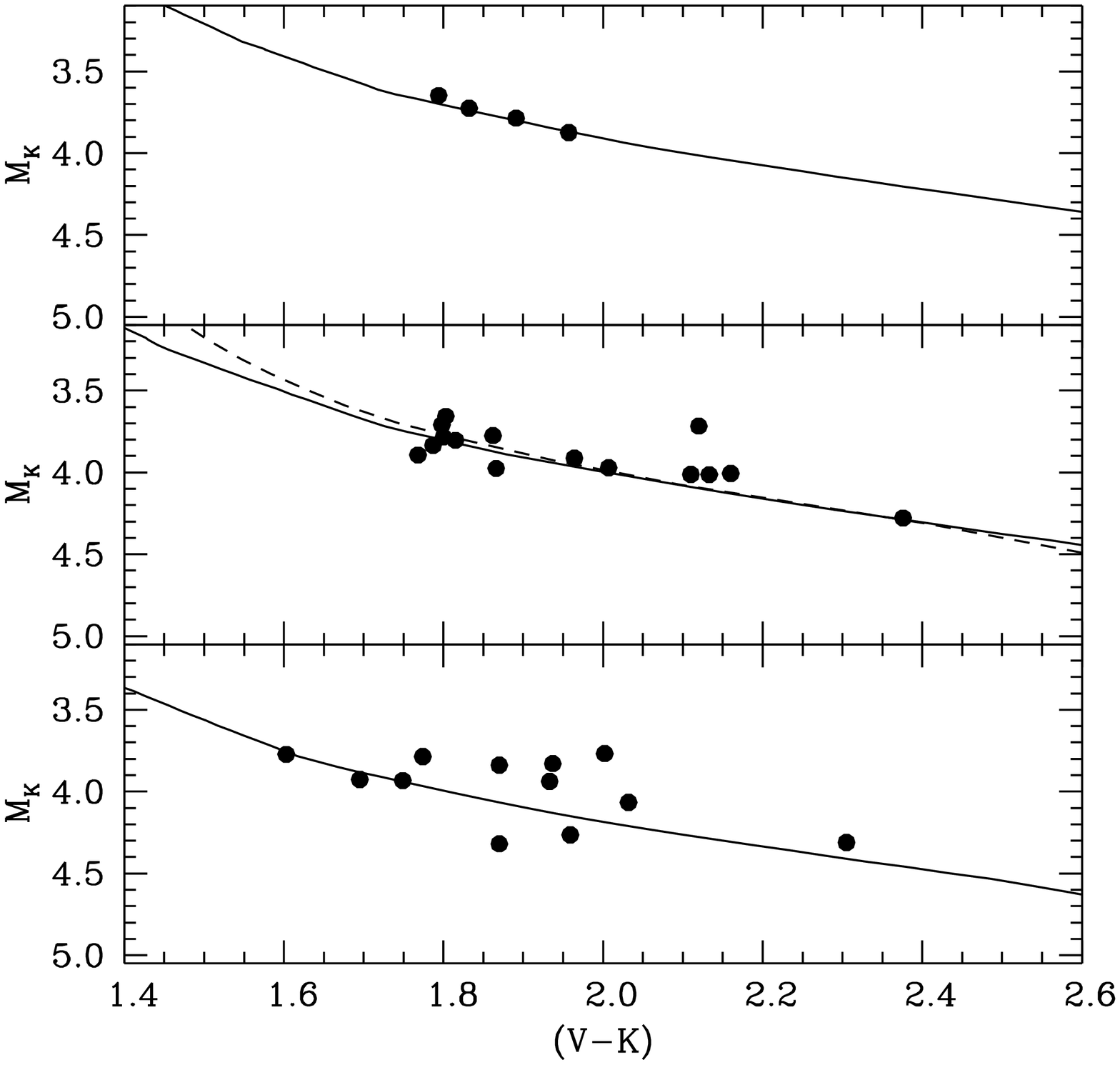}        
\caption{As in Fig.~\ref{subdwbv} but for the $M_K-(V-K)$ CMD.
\label{subdwvk}}        
\end{figure}
\begin{figure}        
\plotone{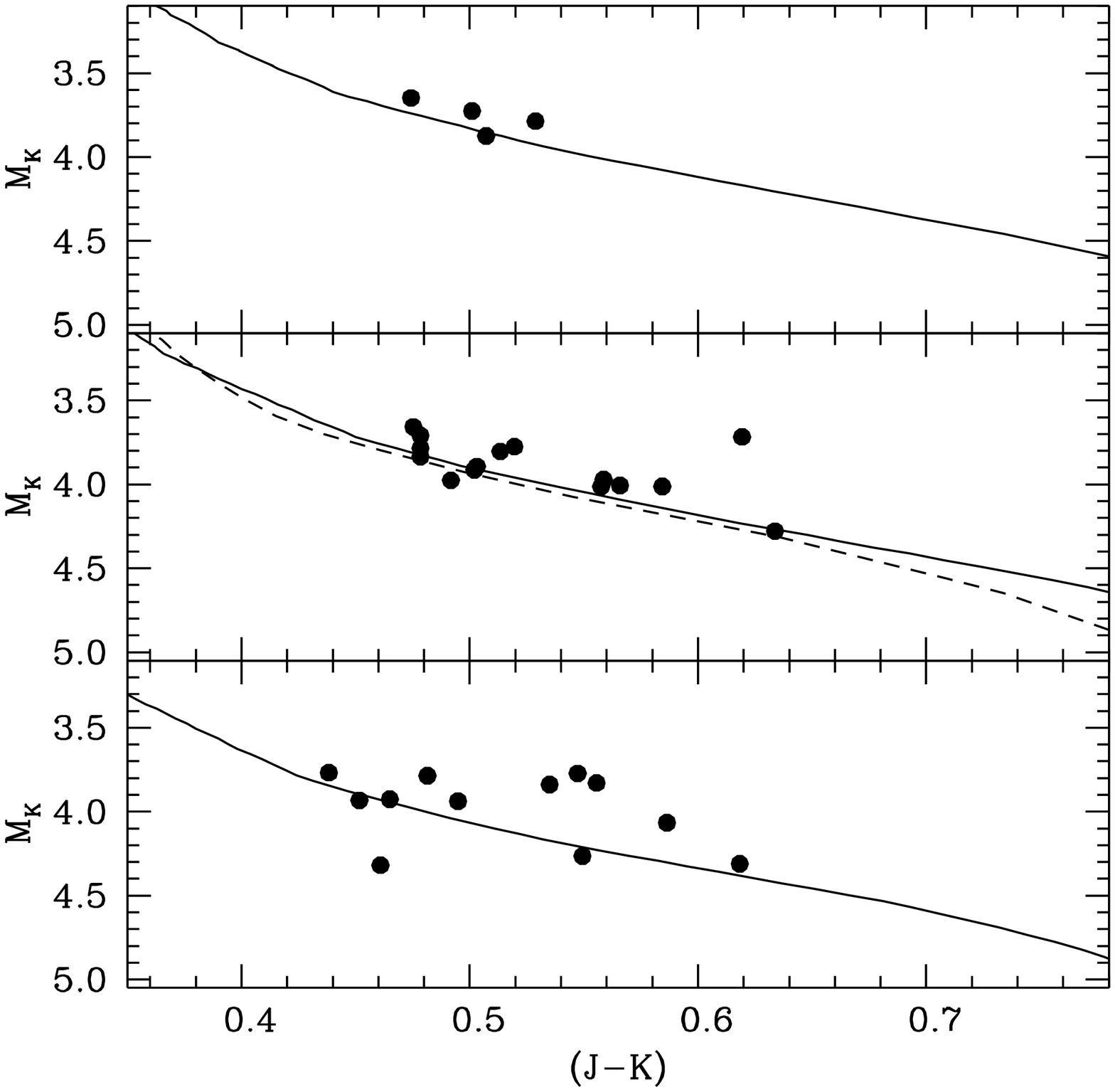}        
\caption{As in Fig.~\ref{subdwbv} but for the $M_K-(J-K)$ CMD.
\label{subdwjk}}        
\end{figure}
\begin{figure}        
\plotone{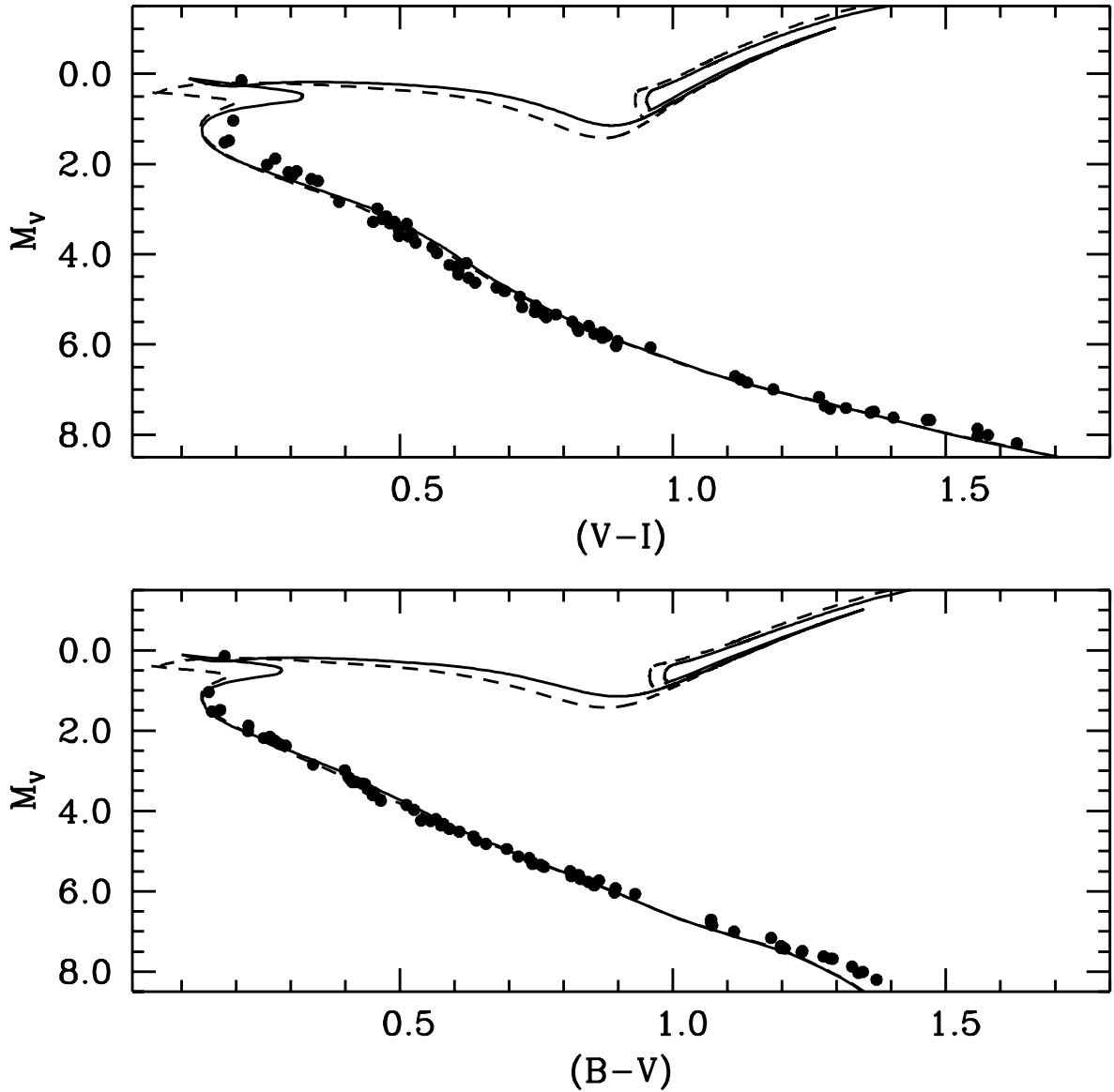}        
\caption{Comparison between our [Fe/H]=0.13, 790~Myr overshooting (solid line)
and 560~Myr canonical (dashed line) isochrones and the Hyades 
$M_V-(B-V)$ and $M_V-(V-I)$ CMDs, corrected for the
parallax distance of the individual stars. 
\label{Hyadesbvi}}        
\end{figure}
\begin{figure}        
\plotone{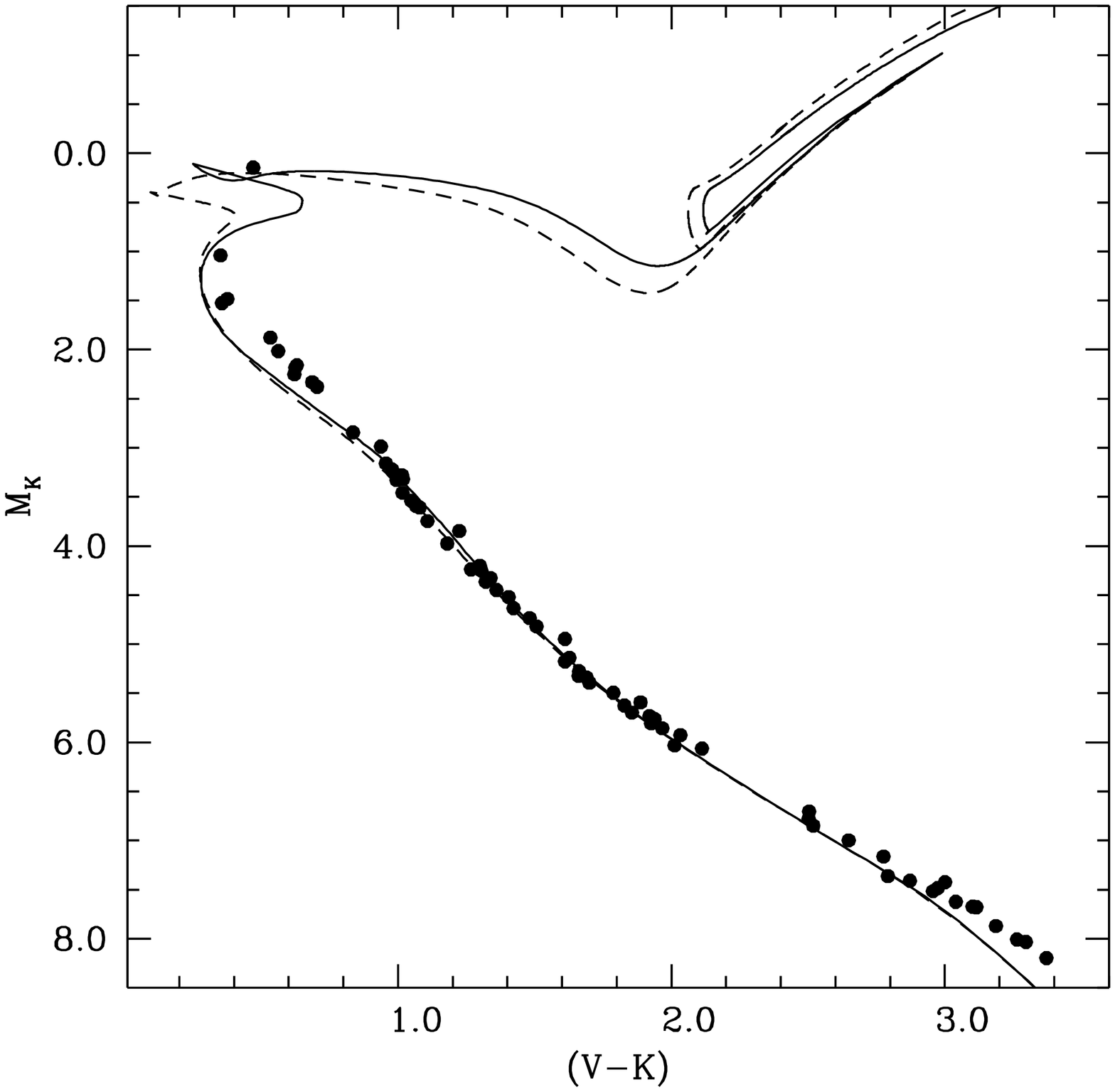}        
\caption{As in Fig.~\ref{Hyadesbvi} but for the $M_K-(V-K)$ CMD.
\label{Hyadesvk}}        
\end{figure}
\begin{figure}        
\plotone{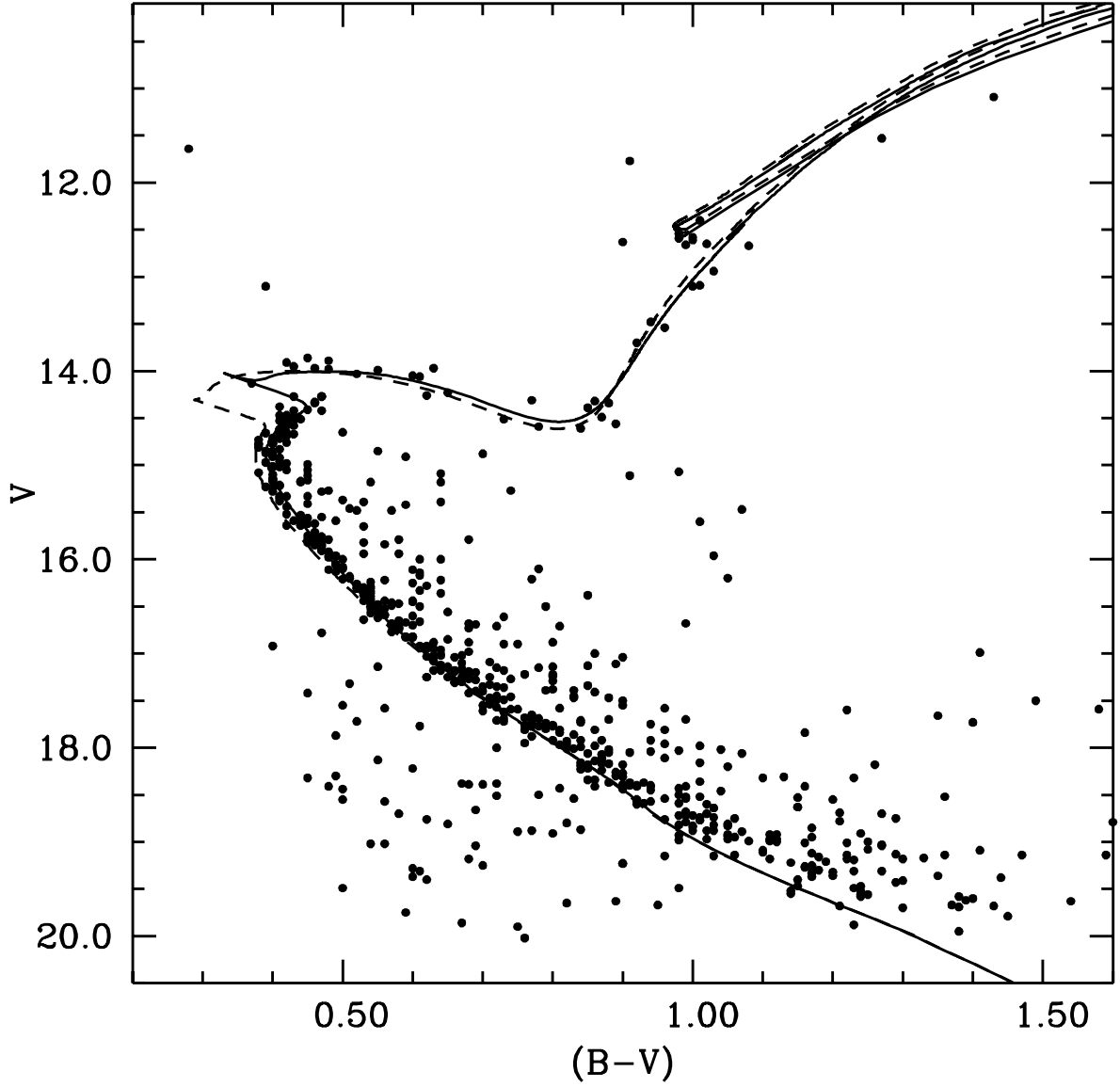}        
\caption{Comparison between our [Fe/H]=$-$0.44, 3.2 Gyr overshooting
(solid line) and 2~Gyr canonical (dashed line) isochrones, and NGC~2420 $V-(B-V)$
CMD. We have shifted the isochrones by E$(B-V)$=0.06 and $(m-M)_0$=11.90.
\label{2420bv}}        
\end{figure}
\begin{figure}        
\plotone{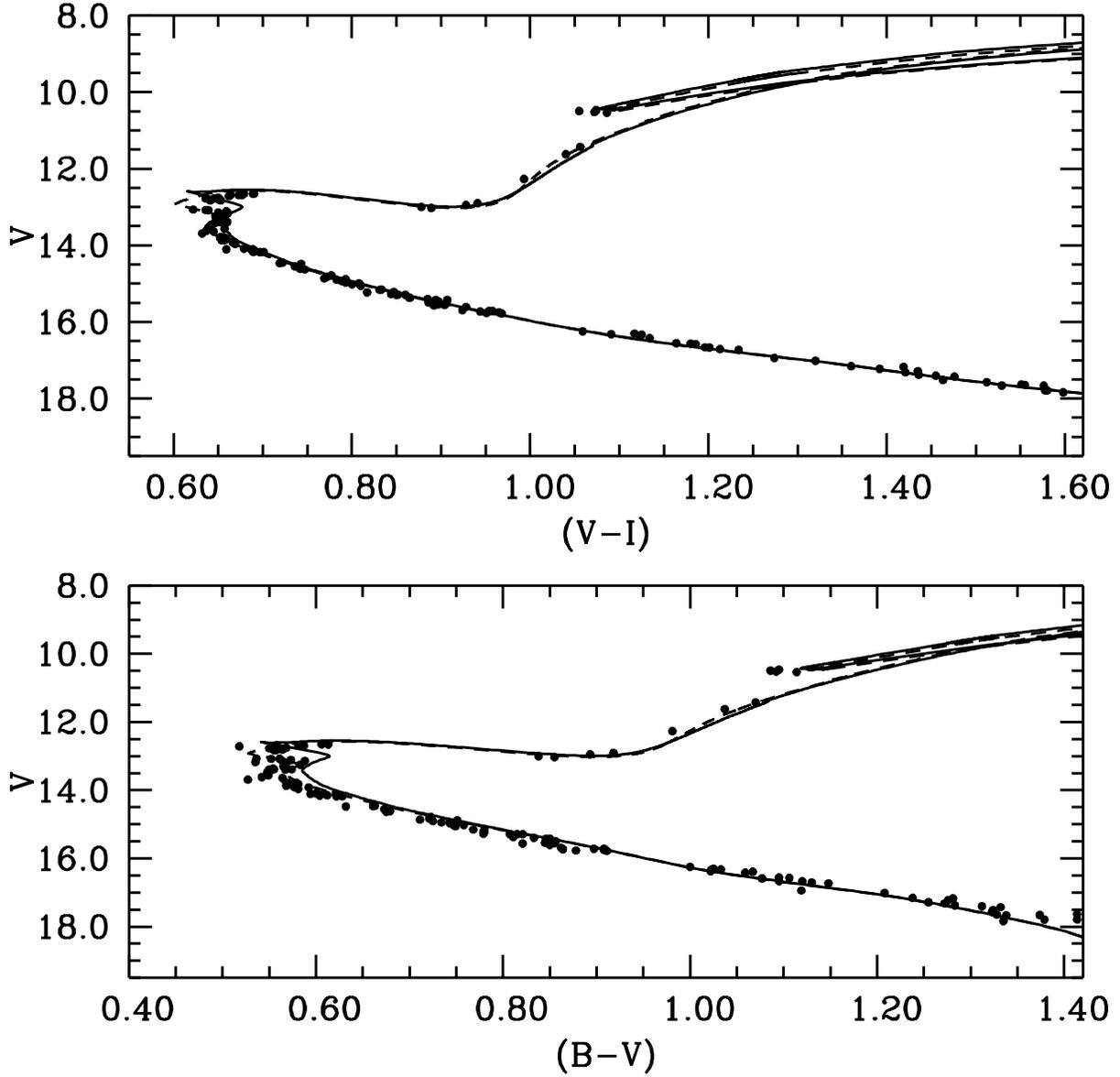}        
\caption{Comparison between our [Fe/H]=0.06, 4.8 Gyr overshooting
(solid line) and 3.8~Gyr canonical (dashed line) isochrones, and M~67 $V-(B-V)$
and $V$-$(V-I)$ CMDs. 
We have shifted the isochrones by E$(B-V)$=0.02 and $(m-M)_0$=9.66.
\label{67bvi}}        
\end{figure}
\begin{figure}        
\plotone{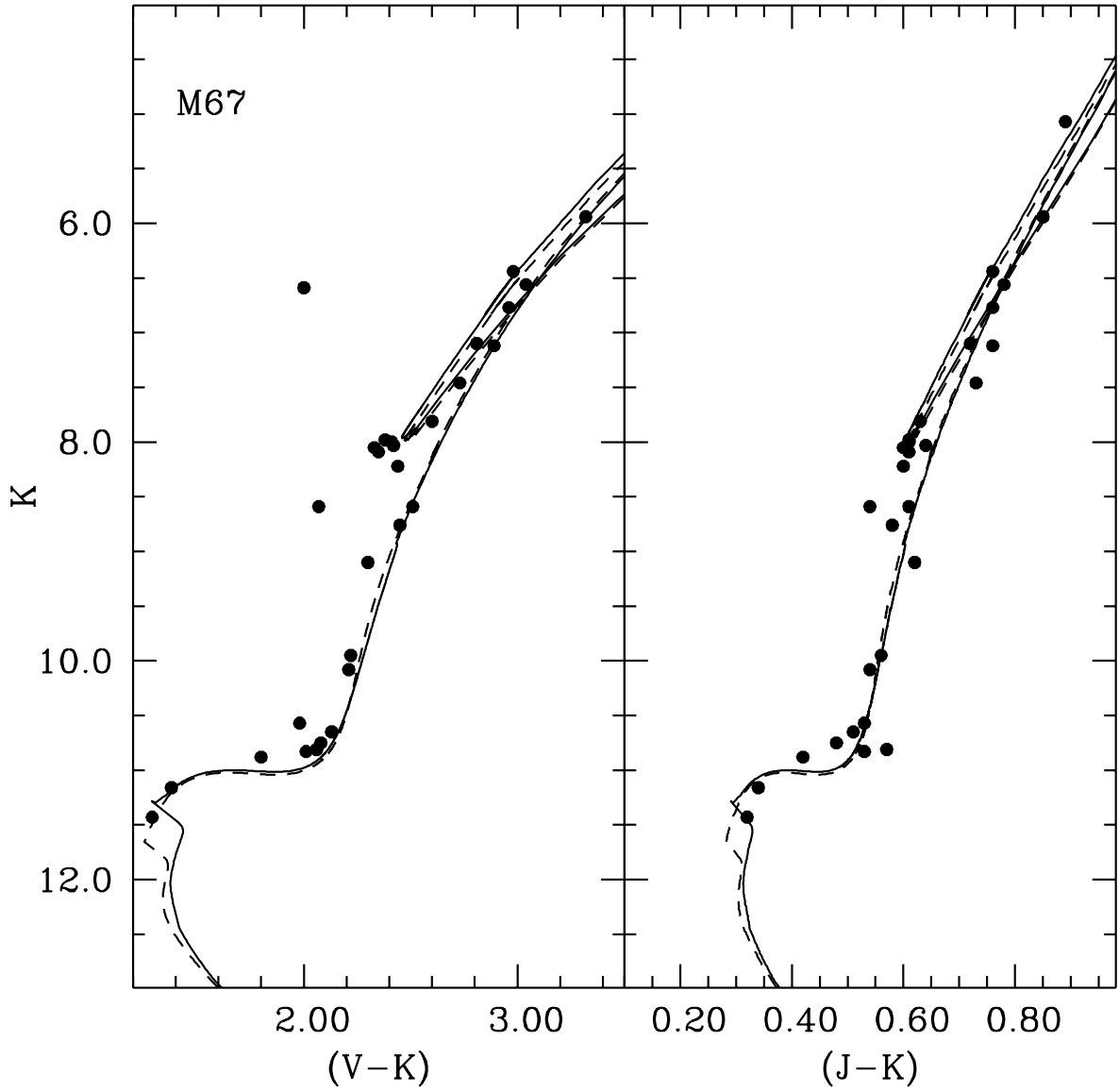}        
\caption{As in Fig.~\ref{67bvi} but for the $K-(V-K)$ and
$K$-$(J-K)$ CMDs
\label{67vjk}}        
\end{figure}
\begin{figure}        
\plotone{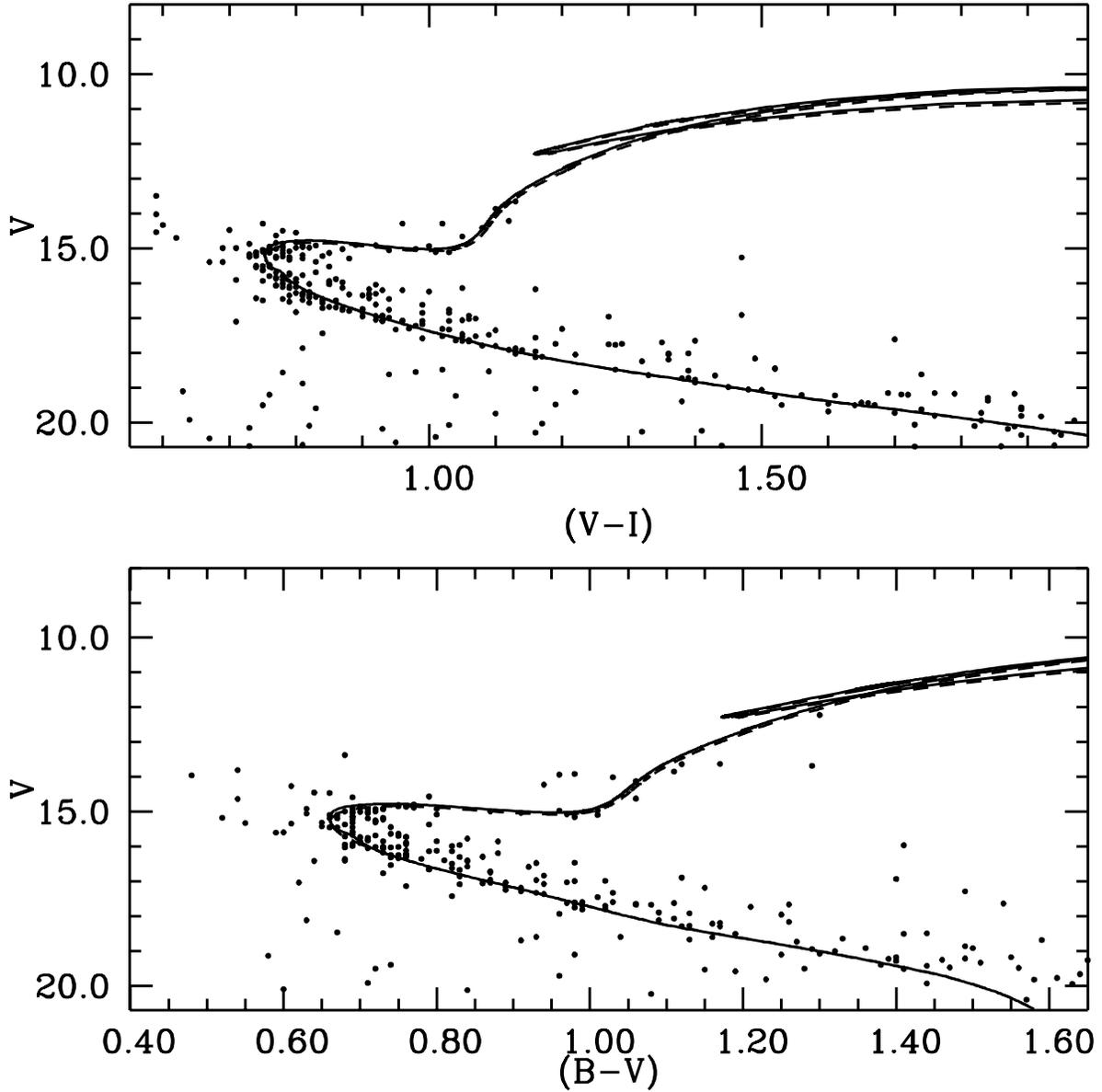}        
\caption{Comparison between our [Fe/H]=$-$0.03, 6.3~Gyr overshooting
(solid line) and canonical (dashed line) isochrones, and NGC~188 $V-(B-V)$
and $V-(V-I)$ CMDs. 
We have shifted the isochrones by E$(B-V)$=0.09 and $(m-M)_0$=11.17.
\label{188bvi}}        
\end{figure}
\begin{figure}        
\plotone{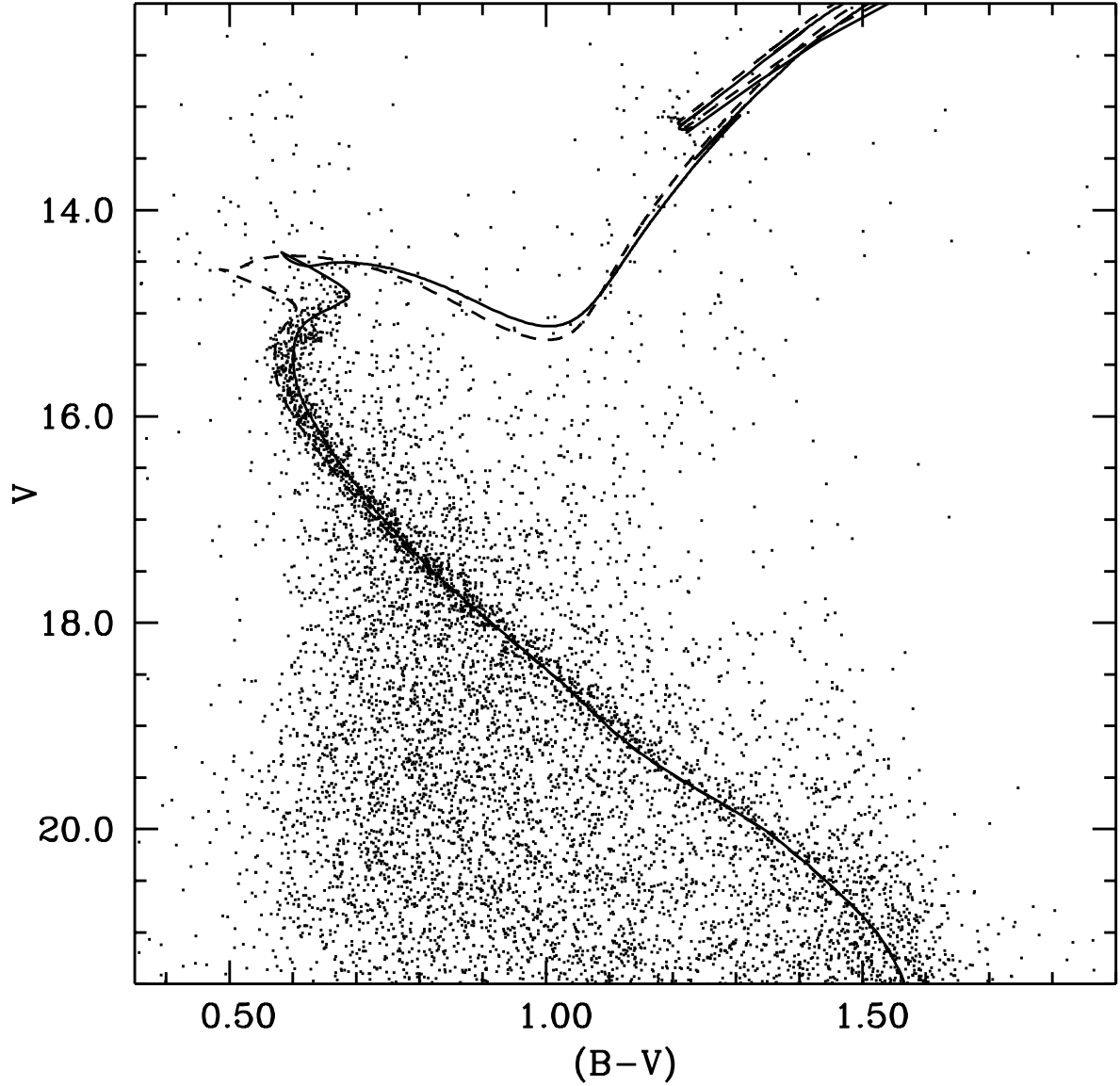}        
\caption{Comparison between our [Fe/H]=0.06, 3~Gyr overshooting
(solid line) and 1.8~Gyr canonical (dashed line) isochrones, and NGC~6819 $V-(B-V)$
CMD. We have shifted the isochrones by E$(B-V)$=0.14 and $(m-M)_0$=12.10.
\label{6819bv}}        
\end{figure}
\begin{figure}        
\plotone{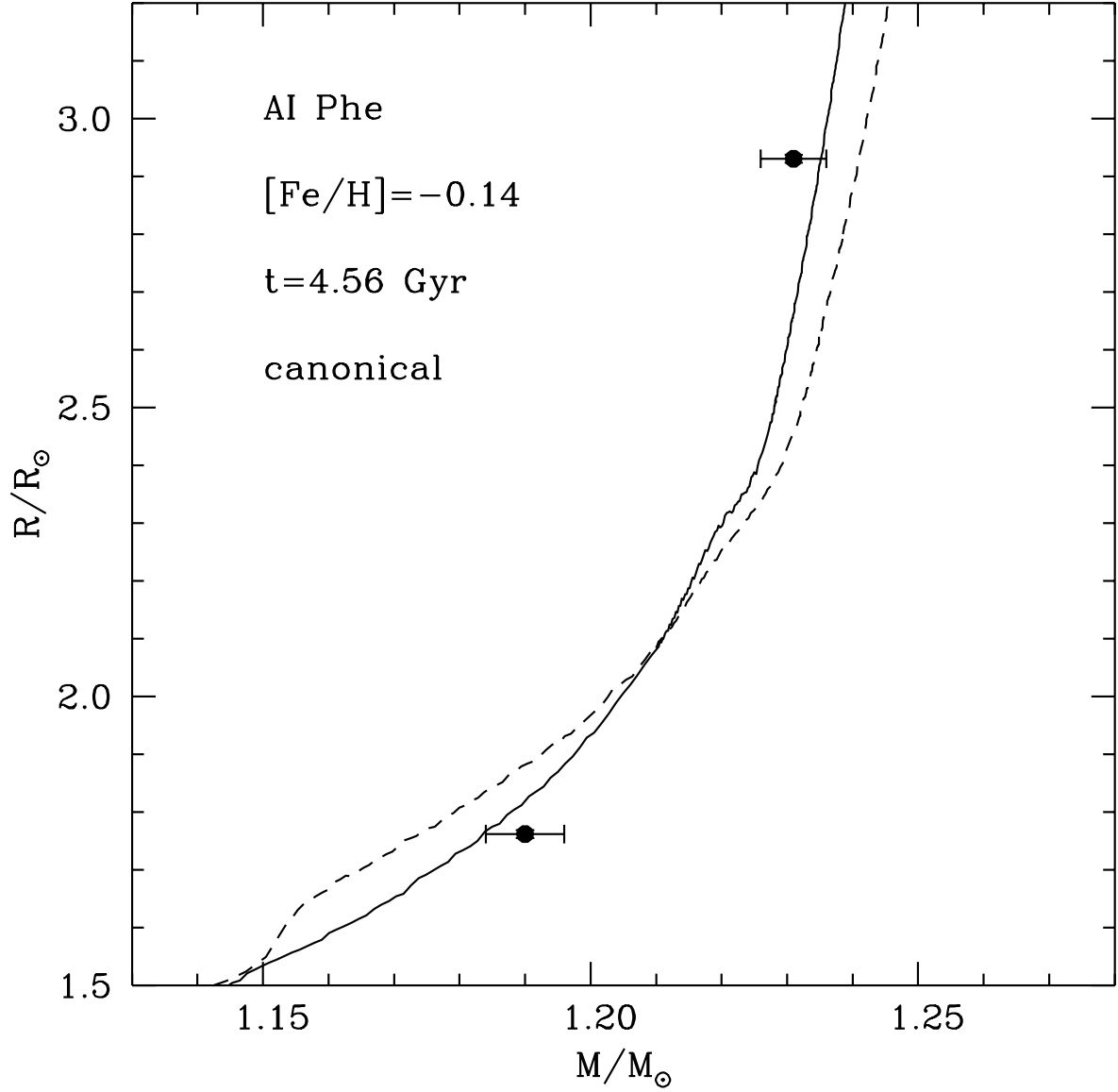}        
\caption{Comparison between the observed masses and radii of the
components of the eclipsing binary AI~Phe, and the predictions of a canonical isochrone
(solid line) with the labeled metallicity and age. The dashed line represents an
overshooting isochrone with the same metallicity and an age of 5.37
Gyr (see text for details).
\label{AIPhe}}        
\end{figure}

\acknowledgments
We warmly thank the editor, Brian Chaboyer, for his handling of the
manuscript. We thank L. Girardi, S. Yi and A. Potekhin for providing data 
in electronic form and for many insightful discussions, J.
Kalirai for providing us with the photometry of NGC~6891, S.
Percival for making available to us the field star data in
machine-readable form.
We wish to warmly thank A. Irwin for all the help provided in using his EOS code as well as 
for many interesting
discussions. A special thank to D. Vandenberg for sharing with us his evolutionary results
in advance of publication. It is a pleasure to thank S. Degl'Innocenti for many useful
discussions on helioseismology and standard solar models, and for
providing the data for Fig.~\ref{sspeed}.
L. Piersanti is warmly acknowledged for the 
valuable suggestions provided during the updating the stellar evolution code.
We are very grateful to M. Castellani for the relevant help
provided when preparing the BASTI WEB pages.
S. C. is grateful for the hospitality at the Instituto de Astrof\'isica de Canarias in
Tenerife. This research has made use of NASA's Astrophysics Data System Abstract
Service and SIMBAD database operated at CDS, Strasbourg, France.
This work was partially supported by MURST (PRIN2002, PRIN2003).

\pagebreak


\clearpage     
\begin{deluxetable}{ccr}      
\tablewidth{0pt}      
\tablecaption{Initial chemical compositions of our model grid.}      
\tablehead{      
\colhead{$Z$}&       
\colhead{$Y$}&       
\colhead{[Fe/H]}}      
  \startdata   
	   0.0001 &       0.245  & $-$2.27\\
	   0.0003 &       0.245  & $-$1.79\\
	   0.0010 &       0.246  & $-$1.27\\
	   0.0020 &       0.248  & $-$0.96\\        
	   0.0040 &       0.251  & $-$0.66\\
           0.0080 &       0.256  & $-$0.35\\
	   0.0100 &       0.259  & $-$0.25\\
	   0.0198 &       0.273  &    0.06\\
	   0.0300 &       0.288  &    0.26\\
	   0.0400 &       0.303  &    0.40\\
\enddata      
\label{chemcomp}
\end{deluxetable}      
      
\clearpage

\clearpage
\begin{deluxetable}{ccc}
\tablewidth{0pt}
\tablecaption{The adopted Key Points (KPs) along the evolutionary tracks (HM
denotes MS stars with convective cores, LM stars with radiative cores
on the MS), and the number of normalized points between consecutive KPs.}
\tablehead{
\colhead{Key point} &
\colhead{Evolutionary phase}&
\colhead{$\rm N^o$ of normalized points}}    
\startdata
 1 & Start of the central H-burning phase                              & 1   \nl
 2 & $1^o$ Minimum in $T_{eff}$ for HM or $X_c=0.30$ for LM                    & 200 \nl
 3 & Maximum in $T_{eff}$ along the Main Sequence                              & 60  \nl
 4 & Maximum in luminosity for HM or $X_c=0.0$ for LM stars                    & 60  \nl
 5 & Minimum in luminosity for HM or base of the RGB for LM                    & 70  \nl
 6 & Tip of the RGB                                                            & 800 \nl
 7 & Start of the quiescent central He-burning phase                                  &  10 \nl
 8 & Central abundance of He equal to 0.55                                     & 150 \nl
 9 & Central abundance of He equal to 0.50                                     & 100 \nl
10 & Central abundance of He equal to 0.40                                     & 100 \nl
11 & Central abundance of He equal to 0.20                                     &  80 \nl 
12 & Central abundance of He equal to 0.10                                     &  80 \nl
13 & Central abundance of He equal to 0.00                                     & 140 \nl
14 & $L_{CNO} > L_{3\alpha}$ during the AGB phase                              & 150 \nl
   & (Carbon ignition for stars with non-degenerate CO cores) &\nl       
     &    &  \nl
 & Horizontal Branch Models & \nl   
     &    &  \nl
   1 & Start of the central He-burning phase              &       1 \nl
   2 & Central abundance of He equal to 0.50              &     130 \nl
   3 & Central abundance of He equal to 0.40              &      50\nl
   4 & Central abundance of He equal to 0.20              &     120 \nl
   5 & Central abundance of He equal to 0.10              &      70\nl
   6 & Central abundance of He equal to 0.00              &      80\nl
   7 & Maximum in luminosity along the clump on the AGB   &     150 \nl
   8 & Minimum in luminosity along the clump on the AGB   &      50\nl
   9 & Maximum in luminosity all along the track          &     250\nl
\enddata 
\label{keypoints} 
\end{deluxetable}
     
\clearpage
\setcounter{page}{40}

\begin{deluxetable}{rrcccccccccccccccccc}
\tabletypesize{\scriptsize}
\setlength{\tabcolsep}{0.024in}
\tablewidth{0pt}
\tablecaption{Selected evolutionary and structural properties for low-
and intermediate-mass models with [Fe/H]=0.06 (no overshooting).}
\tablehead{
\colhead{$M$\tablenotemark{a}} &
\colhead{$\log{L}$\tablenotemark{b}} &
\colhead{$\log{T_{e}}$\tablenotemark{c}} &
\colhead{$M_{cc}$\tablenotemark{d}} &
\colhead{$M_{He}$\tablenotemark{e}} &
\colhead{$t_H$\tablenotemark{f}} &
\colhead{$\log{L}$\tablenotemark{g}} &
\colhead{$\log{T_{e}}$\tablenotemark{h}} &
\colhead{$He_{s}$\tablenotemark{i}} &
\colhead{$Age$\tablenotemark{j}} &
\colhead{$M_{He}$\tablenotemark{k}} &
\colhead{$M_{cc}$\tablenotemark{l}} &
\colhead{$M_{He}$\tablenotemark{m}} &
\colhead{$M_{CO}$\tablenotemark{n}} &
\colhead{$t_{He}$\tablenotemark{o}} &
\colhead{$M_{CO}$\tablenotemark{p}} &
\colhead{$M_{He}$\tablenotemark{q}} &
\colhead{$\log{L}$\tablenotemark{r}} &
\colhead{$\log{T_{e}}$\tablenotemark{s}} &
\colhead{$Age$\tablenotemark{t}} }
\startdata
   .95 & $-$.258 & 3.738 & .034 & .1092 & 14052.1 & 3.472 & 3.464 & .295 & 15301.2 & .4782 & .115 & .4944 & .2116 &  95.4 & .4767 & .5104 & 3.245 & 3.523 & 15413.2\nl
  1.00 & $-$.161 & 3.751 & .053 & .1110 & 11513.3 & 3.475 & 3.463 & .296 & 12644.1 & .4777 & .114 & .4993 & .2168 &  95.0 & .4831 & .5156 & 3.310 & 3.502 & 12754.6\nl
  1.10 &  .018 & 3.771 & .084 & .1177 &  7913.1 & 3.479 & 3.479 & .297 &  8869.1 & .4775 & .112 & .5046 & .1910 &  94.1 & .4885 & .5201 & 3.293 & 3.509 &  8977.6\nl
  1.20 &  .213 & 3.790 & .066 & .1224 &  5640.7 & 3.482 & 3.489 & .295 &  6482.0 & .4774 & .113 & .5075 & .1825 &  92.6 & .4911 & .5220 & 3.319 & 3.513 &  6589.3\nl
  1.30 &  .400 & 3.813 & .015 & .1319 &  4138.6 & 3.484 & 3.498 & .293 &  4880.8 & .4776 & .114 & .5103 & .2057 &  92.0 & .4938 & .5244 & 3.335 & 3.517 &  4991.7\nl
  1.40 &  .556 & 3.831 & .083 & .1401 &  3043.3 & 3.484 & 3.507 & .291 &  3821.2 & .4777 & .117 & .5123 & .2058 &  91.2 & .4977 & .5283 & 3.488 & 3.506 &  3926.6\nl
  1.50 &  .691 & 3.851 & .125 & .1459 &  2312.5 & 3.484 & 3.513 & .289 &  3032.7 & .4778 & .113 & .5139 & .2058 &  90.9 & .4980 & .5277 & 3.346 & 3.532 &  3137.5\nl
  1.60 &  .812 & 3.874 & .159 & .1624 &  1844.3 & 3.485 & 3.520 & .288 &  2433.8 & .4776 & .112 & .5161 & .2057 &  90.2 & .4990 & .5289 & 3.339 & 3.535 &  2538.0\nl
  1.70 &  .923 & 3.897 & .194 & .1706 &  1520.8 & 3.483 & 3.524 & .286 &  1965.5 & .4774 & .114 & .5171 & .2059 &  89.9 & .5016 & .5308 & 3.390 & 3.535 &  2069.4\nl
  1.80 & 1.019 & 3.919 & .226 & .1803 &  1333.3 & 3.475 & 3.530 & .285 &  1599.5 & .4760 & .119 & .5182 & .2059 &  90.7 & .5019 & .5310 & 3.369 & 3.543 &  1703.8\nl
  1.90 & 1.114 & 3.939 & .256 & .1905 &  1142.9 & 3.458 & 3.536 & .284 &  1320.3 & .4726 & .111 & .5167 & .2111 &  92.6 & .4997 & .5297 & 3.339 & 3.550 &  1426.5\nl
  2.00 & 1.207 & 3.956 & .292 & .2016 &   946.5 & 3.420 & 3.544 & .283 &  1106.9 & .4652 & .109 & .5132 & .2115 &  97.1 & .4961 & .5265 & 3.353 & 3.553 &  1217.7\nl
  2.10 & 1.291 & 3.973 & .320 & .2106 &   836.0 & 3.364 & 3.553 & .283 &   943.4 & .4546 & .113 & .5072 & .2071 & 103.5 & .4908 & .5216 & 3.325 & 3.559 &  1061.8\nl
  2.20 & 1.371 & 3.989 & .358 & .2229 &   727.1 & 3.284 & 3.566 & .284 &   814.4 & .4401 & .103 & .4993 & .2002 & 113.8 & .4828 & .5153 & 3.341 & 3.561 &   943.5\nl
  2.30 & 1.448 & 4.003 & .380 & .2366 &   637.8 & 3.133 & 3.583 & .285 &   706.5 & .4164 & .100 & .4869 & .2065 & 134.5 & .4685 & .5038 & 3.220 & 3.577 &   857.7\nl
  2.45 & 1.554 & 4.023 & .422 & .2460 &   534.4 & 2.435 & 3.644 & .287 &   569.5 & .3293 & .035 & .4516 & .2064 & 264.4 & .4403 & .4820 & 3.126 & 3.589 &   859.3\nl
  2.50 & 1.591 & 4.030 & .436 & .2543 &   502.3 & 2.466 & 3.643 & .287 &   531.7 & .3314 & .045 & .4620 & .2762 & 260.8 & .4443 & .4854 & 3.236 & 3.580 &   807.4\nl
  2.60 & 1.657 & 4.042 & .473 & .2664 &   447.9 & 2.451 & 3.647 & .288 &   472.5 & .3373 & .060 & .4672 & .2435 & 228.5 & .4489 & .4888 & 3.115 & 3.594 &   720.1\nl
  2.80 & 1.783 & 4.065 & .528 & .2900 &   360.9 & 2.458 & 3.651 & .289 &   378.1 & .3559 & .076 & .4863 & .2075 & 177.0 & .4701 & .5053 & 3.262 & 3.585 &   573.9\nl
\enddata 
\tiny{
\tablenotetext{a}{Initial mass (in solar units).}
\tablenotetext{b}{Logarithm of the surface luminosity (in solar units) on
the Zero Age Main Sequence (ZAMS).}
\tablenotetext{c}{Logarithm of the effective temperature (in K) on the ZAMS.}
\tablenotetext{d}{Mass of the convective core (in solar units) on the ZAMS.}
\tablenotetext{e}{He core mass (in solar units) at the end of the central H-burning phase.}
\tablenotetext{f}{Central H-burning lifetime (in Myr).}
\tablenotetext{g}{Logarithm of the surface luminosity (in solar units) at the RGB tip.}
\tablenotetext{h}{Logarithm of the effective temperature (in K) at the RGB tip.}
\tablenotetext{i}{Surface He abundance (mass fraction) at the RGB tip.}
\tablenotetext{j}{Model age (in Myr) at the RGB tip.}
\tablenotetext{k}{He core mass (in solar units)  at the beginning of the central He-burning phase.}
\tablenotetext{l}{Convective core mass (in solar mass units) at the beginning of the central He-burning phase.}
\tablenotetext{m}{He core mass (in solar units) at the end of the central He-burning phase.}
\tablenotetext{n}{CO core mass (in solar units) at the end of the central He-burning phase.}
\tablenotetext{o}{Central He-burning lifetime (in Myr).}
\tablenotetext{p}{CO core mass (in solar units) at the $1^o$ thermal pulse.}
\tablenotetext{q}{He core mass (in solar units) at the $1^o$ thermal pulse.}
\tablenotetext{r}{Logarithm of the surface luminosity (in solar units) at the $1^o$ thermal pulse.}
\tablenotetext{s}{Logarithm of the effective temperature (in K) at the $1^o$ thermal pulse.}
\tablenotetext{t}{Model age (in Myr) at the beginning of the thermal pulses phase.}
}
\label{tabevolow} 
\end{deluxetable}

\clearpage

\scriptsize

\clearpage
\begin{deluxetable}{rccccccccccccccccc}
\tabletypesize{\scriptsize}
\setlength{\tabcolsep}{0.024in}
\tablewidth{0pt}
\tablecaption{Selected evolutionary and structural properties for
intermediate-mass and massive models with [Fe/H]=0.06 (no overshooting).}
\tablehead{
\colhead{$M$\tablenotemark{a}} &
\colhead{$\log{L}$\tablenotemark{b}} &
\colhead{$\log{T_{e}}$\tablenotemark{c}} &
\colhead{$M_{cc}$\tablenotemark{d}} &
\colhead{$M_{He}$\tablenotemark{e}} &
\colhead{$t_H$\tablenotemark{f}} &
\colhead{$M_{He}$\tablenotemark{g}} &
\colhead{$M_{cc}$\tablenotemark{h}} &
\colhead{$M_{He}$\tablenotemark{i}} &
\colhead{$M_{CO}$\tablenotemark{j}} &
\colhead{$t_{He}$\tablenotemark{k}} &
\colhead{$M_{CO}$\tablenotemark{l}} &
\colhead{$\log{L}$\tablenotemark{m}} &
\colhead{$M_{CO}$\tablenotemark{n}} &
\colhead{$M_{He}$\tablenotemark{o}} &
\colhead{$\log{L}$\tablenotemark{p}} &
\colhead{$\log{T_{e}}$\tablenotemark{q}} &
\colhead{$Age$\tablenotemark{r}} }
\startdata
  3.0 & 1.898 & 4.086 &  .581 &  .3123 & 296.6 &  .3785 &  .090 &  .5134 &  .2123 & 136.2 &   \multicolumn{2}{c}{no dredge up}  & .4989 & .5285 & 3.377 & 3.579 & 461.5\nl
  3.5 & 2.151 & 4.132 &  .749 &  .3716 & 196.6 &  .4393 &  .109 &  .6022 &  .2386 &  75.6 &   \multicolumn{2}{c}{no dredge up}  & .5949 & .6124 & 3.717 & 3.557 & 287.2\nl
  4.0 & 2.368 & 4.171 &  .905 &  .4322 & 138.1 &  .5035 &  .125 &  .7169 &  .2767 &  45.3 &   \multicolumn{2}{c}{no dredge up}  & .7131 & .7218 & 4.020 & 3.535 & 192.3\nl
  4.5 & 2.558 & 4.204 & 1.068 &  .4908 & 102.0 &  .5712 &  .139 &  .8445 &  .3882 &  29.7 &          .7490 & 4.127              & .7794 & .7852 & 4.154 & 3.531 & 137.7\nl
  5.0 & 2.724 & 4.232 & 1.242 &  .5593 &  78.5 &  .6403 &  .171 &  .9709 &  .4303 &  21.4 &          .7815 & 4.191              & .8036 & .8086 & 4.234 & 3.532 & 104.2\nl
  5.5 & 2.872 & 4.258 & 1.402 &  .6147 &  62.4 &  .7121 &  .199 & 1.1072 &  .5070 &  16.6 &          .8181 & 4.242              & .8311 & .8354 & 4.276 & 3.535 &  82.0\nl
  6.0 & 3.007 & 4.280 & 1.425 &  .6838 &  51.0 &  .7886 &  .221 & 1.2314 &  .5273 &  12.7 &          .8433 & 4.269              & .8539 & .8576 & 4.302 & 3.540 &  65.6\nl
  6.5 & 3.127 & 4.301 & 1.601 &  .7472 &  42.6 &  .8666 &  .255 & 1.3688 &  .6035 &  10.3 &          .8799 & 4.315              & .8887 & .8915 & 4.375 & 3.538 &  54.4\nl
  7.0 & 3.238 & 4.320 & 1.783 &  .8124 &  36.3 &  .9475 &  .277 & 1.5144 &  .5827 &   8.7 &          .9249 & 4.337              & .9366 & .9386 & 4.529 & 3.530 &  46.1\nl
  7.5 & 3.339 & 4.337 & 1.978 &  .9084 &  31.5 & 1.0401 &  .294 & 1.6383 &  .6996 &   7.0 &          .9475 & 4.362              & .9589 & .9605 & 4.643 & 3.522 &  39.4\nl
  8.0 & 3.433 & 4.352 & 2.168 &  .9878 &  27.7 & 1.1337 &  .327 & 1.7828 &  .7936 &   5.9 &          .9806 & 4.275              &   \multicolumn{5}{c}{Carbon ignition}\nl
  8.5 & 3.520 & 4.367 & 2.376 & 1.0969 &  24.6 & 1.2343 &  .373 & 1.9304 &  .8474 &   5.0 &         1.0310 & 4.1886             &   \multicolumn{5}{c}{Carbon ignition}\nl  
  9.0 & 3.601 & 4.380 & 2.593 & 1.1884 &  22.1 & 1.3390 &  .402 & 2.0970 & 1.0264 &   4.5 &   \multicolumn{2}{c}{no dredge up}  &   \multicolumn{5}{c}{Carbon ignition}\nl 
  9.5 & 3.677 & 4.393 & 2.806 & 1.3012 &  20.0 & 1.4588 &  .454 & 2.2517 & 1.1457 &   4.2 &   \multicolumn{2}{c}{no dredge up}  &   \multicolumn{5}{c}{Carbon ignition}\nl 
 10.0 & 3.749 & 4.404 & 3.038 & 1.4394 &  18.3 & 1.5752 &  .491 & 2.3731 & 1.1396 &   3.7 &   \multicolumn{2}{c}{no dredge up}  &   \multicolumn{5}{c}{Carbon ignition}\nl  
\enddata 
\tablenotetext{a}{Initial mass (in solar units).}
\tablenotetext{b}{Logarithm of the surface luminosity (in solar units) on the ZAMS.}
\tablenotetext{c}{Logarithm of the effective temperature (in K) on the ZAMS.}
\tablenotetext{d}{Mass of the convective core (in solar units) on the ZAMS.}
\tablenotetext{e}{He core mass (in solar units) at the end of the central H-burning phase.}
\tablenotetext{f}{Central H-burning lifetime (in Myr).}
\tablenotetext{g}{He core mass (in solar units) at the beginning of the central He-burning phase.}
\tablenotetext{h}{Mass of the convective core (in solar units) at the beginning of the central He-burning phase.}
\tablenotetext{i}{He core mass (in solar units) at the end of the central He-burning phase.}
\tablenotetext{j}{CO core mass (in solar units) at the end of the central He-burning phase.}
\tablenotetext{k}{Central He-burning lifetime (in Myr).}
\tablenotetext{l}{CO core mass (in solar units) at the $2^o$ dredge up.}
\tablenotetext{m}{Logarithm of the surface luminosity (in solar units) at the $2^o$ dredge up.}
\tablenotetext{n}{CO core mass (in solar units) at the $1^o$ thermal pulse.}
\tablenotetext{o}{He core mass (in solar units) at the $1^o$ thermal pulse.}
\tablenotetext{p}{Logarithm of the surface luminosity (in solar units) at the $1^o$ thermal pulse.}
\tablenotetext{q}{Logarithm of the effective temperature (in K) at the $1^o$ thermal pulse.}
\tablenotetext{r}{Model age (in Myr) at the beginning of the thermal pulses phase.}
\label{tabevoint} 
\end{deluxetable}

\clearpage


\clearpage

\clearpage 

\clearpage

\clearpage

\clearpage

\clearpage

\clearpage

\clearpage

\clearpage

\clearpage

\clearpage 

\clearpage

\clearpage 

\clearpage 

\clearpage 

\clearpage 

\clearpage 

\clearpage 

\clearpage       
     
\clearpage 

\clearpage 

\clearpage 

\clearpage 

\clearpage 

\clearpage

 \end{document}